# CRITICALLY EVALUATED SPECTRAL DATA FOR NEUTRAL CARBON (C I)

K. HARIS AND A. KRAMIDA

National Institute of Standards and Technology, Gaithersburg, MD 20899-8422, USA

## ABSRACT

In this critical compilation, all known to us experimental data on the spectrum of neutral carbon were methodically evaluated and supplemented by parametric calculations with Cowan's code. The sources of experimental data vary from laboratory to astrophysical objects and employ different instrumentation from classical grating and Fourier transform spectrometers to precise laser spectroscopy setups and various other modern techniques. This comprehensive evaluation provides accurate atomic data on energy levels and wavelengths (observed and Ritz) with their estimated uncertainties, as well as a uniform description of observed line intensities. In total, 412 energy levels were optimized with the help of selected 1221 best observed lines participating in 1365 transitions in the wavelength region 750 Å to 609.14 μm. In addition, 737 possibly observable transitions are predicted. Critically evaluated transition probabilities for 1616 lines are provided, of which 241 are new. With the accurate energy levels obtained, combined with additional observed data on high Rydberg states, the ionization limit was determined to be 90 820.348(9) $cm^{-1}$ or 11.2602880(11) eV in fair agreement with the previously recommended value, but more accurate.

## 1. INTRODUCTION

The element carbon, which forms remarkably different allotropes, is essential to life and is the fourth most abundant in the universe. In stars, it takes part in the carbon-nitrogen-oxygen (CNO) cycle and is created via the 3$\alpha$-process. It is ubiquitous in the interstellar medium (ISM) in its numerous forms and plays a major role in the evolution of astrophysical objects (Henning & Schnaiter 1998; Evans 2010). Carbon atomic data are vital for (i) solar photospheric determinations of the CNO abundances (Grevesse 1984), (ii) testing of space- and time-variation of the fundamental constants (Berengut & Flambaum 2010; Curran et al. 2011; Levshakov et al. 2012), (iii) determination of both the astrophysical and chemical conditions of atomic species in various ISM objects (Cardelli et al. 1996; Knapp et al. 2000). More specifically, the isotopic ($^{12}$C/$^{13}$C or $^{14}$C) spectral line data are important to derive the isotopic evolution of the universe and improve understanding nucleosynthesis in stars, to elucidate the effects of isotope shifts in the search for variation of fundamental constants (Berengut & Flambaum 2010; Curran et al. 2011; Levshakov et al. 2012; Murphy & Berengut 2014), and to facilitate laser-based mass spectrometry studies (Clark 1983). In laboratory, carbon has an extensive usage record, from historic arc-type discharges to various laser-driven plasma sources. It is the most commonly found impurity in laboratory light sources. Carbon data are of high demand in fusion community (Braams & Chung 2015). Considering the above concerns, consistent and accurate data on wavelengths, intensities, energy levels, ionization energies and transition rates are always of high priority for a wide range of applications from laboratory to astrophysics.

Neutral atomic carbon (C I) contains six electrons arranged as [He]$2s^22p^2$ in the ground electronic configuration with five levels, $^3P_{0,1,2}$, $^1D_2$ and $^1S_0$. Excitation of a single outer electron generates configurations of the type $2s^22pn\ell$ ($n > 2$, $\ell = s, p, d, f, g, h, \ldots$). Excitation from the closed $2s$ core produces the $2s2p^3$ configuration with six terms ($^5S°$, $^3D°$, $^3P°$, $^3S°$, $^1D°$ and $^1P°$) and the $2s2p^2n\ell$ ($n > 2$, $\ell = s, p, d, f, \ldots$) type of configurations. The $2s2p^3$ configuration is partly above the first ionization limit, and $2s2p^2n\ell$ are all autoionizing.

Carbon spectral data have been actively investigated since the early twentieth century. Many discrete lines of atomic carbon were well known from carbon arc studies (see, e.g., Simeon (1923) and references therein). In the early studies, molecular features were dominant in many light sources, obscuring the atomic lines. Merton & Johnson (1923) and Johnson (1925) overcame this hurdle by using a vacuum tube light source (condensed discharge in He) to observe more atomic features of carbon. This method was followed by Fowler & Selwyn (1928) to study the spectrum from the near-infrared (NIR) down to the vacuum ultraviolet (VUV) region. They suggested the wavelength classifications and term values of C I and also interpreted some VUV and visible observations of other authors (Bowen & Ingram 1926; Bowen 1927; Ryde 1927). The identifications were extended to a farther infrared (IR) region by Ingram (1929) and to a deeper VUV range by Paschen & Kruger (1930). A few years later, the first composite analysis by Edlén (1934) combined all previous measurements together with his own VUV studies (Edlén 1933a, 1933b) and brought some consistency in frame of the Ritz combination principle. These studies were elaborated by more accurate measurements of the 2*p*3*s*-2*p*3*p* array in the NIR region (Edlén 1936; Meggers & Humphreys 1933; Kiess 1938; Minnhagen 1954, 1958). Shenstone (1947) established the level value as well as transition identifications for the $2s2p^3\ {}^5S°_2$ state. Prior to his findings, it was believed impossible to observe transitions from this level in laboratory light sources, because in pure *LS* coupling they are forbidden.

Carbon lines were proposed as auxiliary wavelength standards in the VUV region (Boyce & Robinson 1936; More & Rieke 1936). To this effect, the most reliable measurements were made by Wilkinson (1955), Herzberg (1958), Wilkinson & Andrew (1963) and Kaufman & Ward (1966). In 1965, Johansson and Litzén (1965) undertook Fabry-Perot interferometric measurements of C I in the far IR region (3868–8604) $cm^{-1}$ with an uncertainty of ~0.02 $cm^{-1}$.

Apart from many *LS*-allowed and intercombination transitions between the 2*p*3*p* and (2*p*3*d* + 2*p*4s + 2*s*2*p*$^3$) configurations, they also observed the 2*p*3*d*–2*p*4*f* transitions. Johansson (1966) made another set of comprehensive measurements using a 21 ft grating spectrograph in the wavelength range of (2478–11 331) Å and thereby determined more accurate energy levels with uncertainties less than 0.05 cm$^{-1}$. He observed many new transitions in the 2*p*3*s*–2*pnp* (*n* = 3–11), 2*p*3*p*–2*pnℓ* (*n* = 5–10 for *ℓ* = *s*; *n* = 4–11 for *ℓ* = *d*) and 2*s*2*p*$^3$–2*pnℓ* (*n* = 4–8, *ℓ* = *p*, *f*) arrays. These observations improved upon the accuracy of most of the earlier VUV measurements (Paschen & Kruger 1930; Wilkinson 1955; Wilkinson & Andrew 1963) except those by Herzberg (1958) and Kaufman & Ward (1966).

The next comprehensive compilation of energy levels was carried out by Moore in 1970 (Moore 1970, 1993). Its results were tabulated in her *Atomic Energy Levels* (AEL) book (Moore 1970), which is hereafter referred to as Moore's table. Her work was largely based on Johansson (1966), but the level values were slightly improved by taking into account a few good VUV measurements by Kaufman & Ward (1966), and intensities of lines in VUV region were from photographic examinations (Junkes et al. 1965). Moore's table includes more than 300 Ritz wavelengths of VUV lines, out of which 190 were previously observed. Further VUV linelist enhancements were due to Feldman et al. (1976), who reported Skylab spectra during solar flare activity, and from the flash-pyrolysis photoabsorption spectrum of carbon taken by Mazzoni et al. (1981). In these two studies, extensions of various 2*pns* and 2*pnd* $^{M}L°_{J}$ series were made, and more than 50 energy levels were newly established. The high-resolution solar atlas in the (1175–1710) Å region prepared by Sandlin et al. (1986) contains about 192 neutral carbon lines, out of which 46 were new, and 103 were more accurate than those from other solar atlases published later (Feldman

& Doschek 1991; Curdt et al. 2001; Parenti et al. 2005). Those later atlases also contain some new or improved wavelengths of C I.

Chang & Geller (1998) extended the analysis by combining all available IR data from four different solar spectra (Farmer & Norton 1989; Toon 1991; Livingston & Wallace 1991; Wallace et al. 1993, 1996) together with VUV data of both laboratory (Herzberg 1958; Wilkinson & Andrew 1963; Kaufman & Ward 1966) and solar origin (Sandlin et al. 1986; Feldman et al. 1976). This procedure enabled them to determine the upper energy levels with an average uncertainty of ~0.014 $cm^{-1}$. Laboratory improvement of some of these data was brought up by Wallace & Hinkle (2007) via Fourier transform spectra in a wide range of wavelengths from IR to visible. Their method of level optimization and line identifications was limited by their use of Ritz wavenumbers from Chang & Geller (1998), which were correct in most, but not all cases. Another set of transitions of interest are parity-forbidden ones within the levels of the $2s^22p^2$ ground configuration. The search for them was initiated in the 1920s (Bowen & Ingram 1926). The $^3P_1$–$^1S_0$ transition was subsequently observed for the first time by Boyce (1936) in spectra of many stellar novae. Several other forbidden lines were also observed in the NIR region in different nebulae (Lambert & Swings 1967; Swensson 1967; Liu et al. 1995). The transitions $^3P_0$–$^3P_1$ at 492 GHz and $^3P_1$–$^3P_2$ at 809 GHz (see figure 1) in both stable isotopes $^{12}$C and $^{13}$C were measured with an unprecedented accuracy by several teams (Saykally & Evenson 1980; Cooksy et al. 1986; Yamamoto & Saito 1991; Klein et al. 1998). These lines of high astrophysical interest are important for studying astrochemistry of carbon involving photo-destruction of the CO molecule in the ISM (Langer 2009). They were observed in several astrophysical objects (Phillips et al. 1980; Jaffe et al. 1985; Genzel et al. 1988; Frerking et al. 1989; Keene et al. 1998). In 2005, Labazan et al. (2005) measured all three $2s^22p^2$ $^3P_{0,1,2}$ → $2s2p^3$ $^3S°_1$ transitions (near 945 Å) with an extremely high accuracy in both stable isotopes. Isotopic shifts were measured for a few more transitions of neutral carbon (Burnett 1950; Holmes 1951; Bernheim & Kittrell 1980). Using an atomic-beam+magnetic-resonance technique, Wolber et al. (1970) measured the hyperfine (HF) level separations and $g_J$-factors of the $^{12}$C and $^{13}$C $2s^22p^2$ $^3P$ multiplet. They also improved the values of the nuclear moment and HF separations of the $^3P_{1,2}$ states of the unstable $^{11}$C isotope (having a lifetime of 20.4 min), previously determined by (Haberstroh et al. 1964).

C I has received much attention from the theoretical community. Most theoretical studies reporting transition parameters (*f*- or *A*-values), Stark shift and broadening parameters, and isotopic shifts, are beyond the scope of the present work. A few exceptions are noted. One of them is a large set of critically evaluated transition rate data (Wiese et al. 1996; Wiese & Fuhr 2007), which has been used and extended in the present work.

The present data on C I in the Atomic Spectra Database (ASD) of the National Institute of Standards and Technology (NIST) (see Kramida et al. 2016) are mainly based on the AEL compilation (Moore 1970, 1993). With the growing interest in atomic data, from the perspectives of its various users and numerous applications, the requirement of data reliability and precision is now of the highest priority. In this regard, we note that there exists some disagreement in published fragmented data (Chang & Geller 1998; Wallace & Hinkle 2007). In particular, their stated uncertainties (wavelengths/wavenumbers), were found to be inconsistent with a comprehensive level optimization. Another major problem is related to relative intensities of transition lines, as they are reported from various measurements made at different experimental conditions. The main aim of the present work is to compile and disseminate a comprehensive and internally consistent set of critically evaluated atomic data on energy levels and observed wavelengths with their uncertainties, as well as uniformly-scaled relative line intensities and Ritz wavelengths suitable as secondary standards for the spectrum of neutral carbon.

## 2. METHOD OF EVALUATION OF WAVELENGTHS

In neutral carbon, the only VUV transitions observable in emission from non-autoionizing levels are those of the $2s^22p^2 – [2s2p^3+2s^22pn\ell$ ($n \geq 3$, $\ell = s, d$)] arrays, and no other VUV lines occur between the bound states. In such a spectrum, implementation of the Ritz combination principle allows determining the upper energy levels with much greater accuracy than it would be possible from measurements of only the VUV transitions. The greater accuracy is achieved by measuring successively each step in the ladder of the $2p3s$+$2s2p^3$ → $2p3p$ → $2pn\ell$ ($n \geq 3$, $\ell = d, s$) transitions, all of which lie above 3279 Å (the ionization threshold from the $2p3s$ $^3P°_0$ state) and summing up the wavenumbers of each step of this ladder and the wavenumbers of transitions from the first step down to the ground level. The accuracy of levels established in this way is limited by the accuracy of the known transitions between the ground level and $2p3s$. As briefly described in the Introduction, the initial implementation of this procedure was made by Johansson (1966) by combining his grating measurements in a wide optical range with IR Fabry-Perot measurements of Johansson & Litzén (1965). The level accuracy achieved was considered good at the time, but the advent of Fourier transform (FT) spectrometers (FTS) opened the way for further improvement. Another possibility of improvement appeared after the launch of several missions carrying either high-altitude terrestrial or space-borne high-resolution FT spectrometers covering the entire IR solar spectrum, starting in late 1980s (Farmer & Norton 1989; Toon 1991; Livingston & Wallace 1991; Wallace et al. 1993, 1996). Later, the results of these missions were used in the analysis of C I by Chang & Geller (1998). The major problem with solar data was a high level of spectroscopic contamination. This problem was addressed by Wallace & Hinkle (2007) who solved it by replacing the measurements of Chang & Geller (1998) with measurements of laboratory FTS recordings that were already available in the National Solar Observatory (NSO) archives, but were not analyzed until 2007. The starting point of our investigation was to check the consistency of the measured wavenumbers of Wallace & Hinkle (2007) with their quoted uncertainties in a careful level optimization procedure described by Kramida (2013a). This procedure was successfully implemented in spectra of many atoms/ions in the past (see, e.g., Kramida (2013b, 2013c), Haris et al. 2014). A satisfactory optimization was achieved by adjusting the initial measurement uncertainties in such a way that observed wavenumbers agreed with the calculated ones within the adjusted uncertainties. Many new accurate Ritz wavelengths were found in this procedure. These high-precision Ritz-type standards further served as internal references to recalibrate other, less accurate measurements, which subsequently were inserted in the level optimization. On average, the Ritz values are better by a factor of two or more than most direct measurements. Comparison of some sets of measured wavelengths with Ritz-type standards showed noticeable systematic shifts, which needed to be removed before assessing the statistical uncertainties. The entire procedure involved several iterations of level optimization (see Section 3). Consequently, the best available observed wavelengths were kept in the resulting linelist reported in Table 1. All optimized observed energy levels for this spectrum are summarized in Table 2.

## Table 1

Observed and Predicted Spectral Lines of C I

| Intensity[a] (arb. u.) | $\lambda_{obs}$[b] (Å) | Unc.[b] (Å) | Lower Level[e] | | | Upper Level[e] | | | $\lambda_{Ritz}$[b] (Å) | Unc.[b] (Å) | $A$ ($s^{-1}$) | Acc.[g] | Type[h] | TP Ref.[i] | Line Ref.[j] | Comment[l] |
|---|---|---|---|---|---|---|---|---|---|---|---|---|---|---|---|---|
| 1 | 750.680 | 0.010 | $2s^22p^2$ | $^3P$ | 2 | $2s2p^2(^4P)14p$ | $^3D°$ | 3 | 750.680 | 0.010 | | | | | C81 | GS |
| 1 | 751.240 | 0.010 | $2s^22p^2$ | $^3P$ | 2 | $2s2p^2(^4P)13p$ | $^3D°$ | 3 | 751.240 | 0.010 | | | | | C81 | GS |
| 1 | 751.420 | 0.010 | $2s^22p^2$ | $^3P$ | 1 | $2s2p^2(^4P)13p$ | $^3D°$ | 1 | 751.420 | 0.010 | | | | | C81 | GS |
| … | … | … | … | … | … | … | … | … | … | … | … | … | … | … | … | … |
| 7800 | 945.18746 | 0.00006 | $2s^22p^2$ | $^3P$ | 0 | $2s2p^3$ | $^3S°$ | 1 | 945.18745 | 0.00004 | 3.8e+08 | C+ | | L89a | L05 | |
| 17000 | 945.33418 | 0.00006 | $2s^22p^2$ | $^3P$ | 1 | $2s2p^3$ | $^3S°$ | 1 | 945.33414 | 0.00003 | 1.14e+09 | C+ | | L89a | L05 | U |
| 18000 | 945.57546 | 0.00006 | $2s^22p^2$ | $^3P$ | 2 | $2s2p^3$ | $^3S°$ | 1 | 945.57546 | 0.00003 | 1.9e+09 | C+ | | L89a | L05 | U |
| … | … | … | … | … | … | … | … | … | … | … | … | … | … | … | … | … |
| 730000bl | 1194.028 | 0.014 | $2s^22p^2$ | $^3P$ | 0 | $2s^22p5s$ | $^3P°$ | 1 | 1193.995082 | 0.000020 | 1.94e+07 | A | | T01 | M81 | G |
| 3000000 | 1194.065 | 0.003 | $2s^22p^2$ | $^3P$ | 2 | $2s^22p5s$ | $^3P°$ | 2 | 1194.063006 | 0.000019 | 2.98e+07 | B+ | | T01 | S86 | GU |
| 1100000* | 1194.220 | 0.010 | $2s^22p^2$ | $^3P$ | 1 | $2s^22p5s$ | $^3P°$ | 1 | 1194.229168 | 0.000020 | 8.3e+06 | A | | T01 | S86 | G |
| 1100000* | 1194.300 | 0.010 | $2s^22p^2$ | $^3P$ | 1 | $2s^22p4d$ | $^3F°$ | 2 | 1194.300572 | 0.000019 | 4.1e+06 | B+ | | T01 | S86 | G |
| … | … | … | … | … | … | … | … | … | … | … | … | … | … | … | … | … |
| 31000q | 4762.5252 | 0.0005 | $2s^22p3s$ | $^3P°$ | 1 | $2s^22p4p$ | $^3P$ | 2 | 4762.52473 | 0.00006 | 2.7e+05 | C | | H93a | W07#2 | FD |
| 22000q | 4766.6698 | 0.0014 | $2s^22p3s$ | $^3P°$ | 1 | $2s^22p4p$ | $^3P$ | 1 | 4766.66760 | 0.00007 | 2.4e+05 | C | | H93a | W07#2 | FC |
| 24000 | 4770.02392 | 0.00023 | $2s^22p3s$ | $^3P°$ | 1 | $2s^22p4p$ | $^3P$ | 0 | 4770.02376 | 0.00008 | 1.1e+06 | C | | H93a | W07#2 | F |
| 69000 | 4771.73346 | 0.00016 | $2s^22p3s$ | $^3P°$ | 2 | $2s^22p4p$ | $^3P$ | 2 | 4771.73374 | 0.00006 | 8.0e+05 | C | | H93a | W07#2 | F |
| 34000 | 4775.907 | 0.014 | $2s^22p3s$ | $^3P°$ | 2 | $2s^22p4p$ | $^3P$ | 1 | 4775.89266 | 0.00007 | 4.8e+05 | C | | H93a | J66 | G |
| … | … | … | … | … | … | … | … | … | … | … | … | … | … | … | … | … |
| | | | $2s^22p^2$ | $^3P$ | 0 | $2s^22p^2$ | $^1D$ | 2 | 9808.295 | 0.003 | 5.9e-08 | C | E2 | F06 | TW | P |
| | 9824.31 | 0.19 | $2s^22p^2$ | $^3P$ | 1 | $2s^22p^2$ | $^1D$ | 2 | 9824.118 | 0.003 | 7.3e-05 | C+ | M1 | F06 | L95 | GV |
| | 9850.34 | 0.09 | $2s^22p^2$ | $^3P$ | 2 | $2s^22p^2$ | $^1D$ | 2 | 9850.250 | 0.003 | 2.2e-04 | C+ | M1 | F06 | L95 | GV |
| … | … | … | … | … | … | … | … | … | … | … | … | … | … | … | … | … |
| 32000 | 15727.3506 | 0.0020 | $2s^22p3d$ | $^1D°$ | 2 | $2s^22p(^2P°_{3/2})4f$ | $^2[3/2]$ | 2 | 15727.3529 | 0.0008 | 1.1e+06 | C | | H93 | W07#13 | FD |
| 31000 | 15784.4994 | 0.0017 | $2s^22p3d$ | $^1D°$ | 2 | $2s^22p(^2P°_{3/2})4f$ | $^2[5/2]$ | 2 | 15784.5012 | 0.0007 | 1.4e+06 | C | | H93 | W07#13 | F |
| 32000 | 15784.888 | 0.003 | $2s^22p3d$ | $^1D°$ | 2 | $2s^22p(^2P°_{3/2})4f$ | $^2[5/2]$ | 3 | 15784.8896 | 0.0005 | 7.1e+05 | C | | H93 | W07#13 | F |
| 57000q | 15852.6055 | 0.0018 | $2s^22p3d$ | $^1D°$ | 2 | $2s^22p(^2P°_{3/2})4f$ | $^2[7/2]$ | 3 | 15852.6037 | 0.0005 | 2.1e+06 | C | | H93 | W07#13 | FD |
| … | … | … | … | … | … | … | … | … | … | … | … | … | … | … | … | … |
| 2000 | 49825.54 | 0.025 | $2s^22p3d$ | $^3P°$ | 1 | $2s^22p4p$ | $^3P$ | 1 | 49825.574 | 0.006 | 1.77e+05 | B | | H93a | W07#13* | F |
| 2600 | 49934.51 | 0.03 | $2s^22p3d$ | $^3P°$ | 0 | $2s^22p4p$ | $^3P$ | 1 | 49934.524 | 0.009 | 2.03e+05 | B | | H93a | W07#13 | F |
| 1400 | 50194.62 | 0.03 | $2s^22p3d$ | $^3P°$ | 1 | $2s^22p4p$ | $^3P$ | 0 | 50194.627 | 0.008 | 5.7e+05 | B | | H93a | W07#13 | FU |
| … | … | … | … | … | … | … | … | … | … | … | … | … | … | … | … | … |
| | 6091353.36 | 0.24 | $2s^22p^2$ | $^3P$ | 0 | $2s^22p^2$ | $^3P$ | 1 | 6091353.37 | 0.24 | 7.93e-08 | A | M1 | F06 | Y91 | V |

**Notes**

[a] Averaged relative observed intensities in arbitrary units are given on a uniform scale corresponding to a Boltzmann populations in a plasma with an effective excitation temperature of 0.41 eV, corresponding to the FT spectrum “85R13”. The intensity value is followed by the line character encoded as follows: bl – blended by other lines either specified by an elemental symbol or given by an index in parentheses. The index is explained as follows (unit of values is $cm^{-1}$): O IV/2 – second order of an O IV line, T – contaminated by a telluric line, 1 – 23 418.059, 2 – 20 992.2792, 3 – 8254.2325, 4 – 6834.1017, 5 – 5657.1101; D – double line; d – diffuse; H – very hazy; i – identification uncertain; m – masked by other lines either marked or specified by an index in parentheses. The index is explained as follows (unit of values is $cm^{-1}$): 1– 21 899.0959, 2 – 2927.0767, 3 – 2107.4239, 4 = 1350.858, 5 = 1355.422, 6 = 1349.731, 7 = 1347.773, 8 = 1339.013; ℓ– shaded to long wavelength; p – perturbed by nearby lines either indicated by the spectrum symbol or given by an index in the parenthesis. The index is explained as follows (unit of values is $cm^{-1}$): gh – grating

ghost, 1 – 8191.0769, 2 – 8104.4249, 3 – 6764.1865, 4 – 6740.0118, 5 – 5396.8230, 6 – 3889.1307, 7 – 2033.1415, 8 – 2015.0026; q – asymmetric; r – Easily reversed; sh – Shoulder; w – wide; * – intensity is shared by two or more lines; : – wavelength not measured (the value given is a rounded Ritz wavelength); ? – the given character is uncertain.

[b] Observed and Ritz wavelengths are in vacuum for $\lambda < 2000$ Å and $\lambda > 20\,000$ Å and in standard air for 2000 Å < λ < 20 000 Å. Conversion between air and vacuum was made with the five-parameter formula from Peck & Reeder (1972). Assigned uncertainty of given observed wavelength or computed uncertainty of Ritz wavelength determined in the level optimization procedure.

[c] Signal-to-noise ratio and full width at half maximum (in units of $10^{-3}$ $cm^{-1}$) for the lines measured in FT spectra.

[d] Observed wavenumber (in vacuum) and its uncertainty are given in additional columns in the complete table available in the online journal.

[e] Level designation from Table 2.

[f] Level energy value from Table 2.

[g] Accuracy code of the *A*-value is given in Table 10.

[h] Blank – electric-dipole (E1) transition; M1 – magnetic-dipole transition; E2 – electric-quadrupole transition.

[i] Transition probability references. All transition probabilities, except marked as TW ("This work") are those critically evaluated by Wiese et al. (1996) and Wiese & Fuhr (2007) where the original sources of data were encoded as follows: F06 – Fischer et al. (2006); G89a– normalized to a different scale from values reported by Goldbach et al. (1989); H93 – Hibbert et al. (1993); H93a – normalized to a different scale from values reported by Hibbert et al. (1993); L89 – Luo & Pradhan (1989); L89a – calculated from the multiplet value given by Luo & Pradhan (1989) assuming pure LS-coupling; N84 – Nussbaumer & Storey (1984); N84a – normalized to a different scale from values reported by Nussbaumer & Storey (1984); T01 – Tachiev & Fischer (2001); W – Weiss, private communication, as quoted in Wiese et al. (1996); and TW – This work, semiempirical calculations using Cowan's codes (see text).

[j] Line references: B80 – Bernheim & Kittrell (1980); C81 – Cantù et al. (1981) ;C98 – Chang & Geller (1998); various solar spectra are designated as #1 – NOAO1, Livingston & Wallace (1991), #2 – NOAO2, Wallace et al. (1993), #A – ATMOS, Farmer & Norton (1989), #M – Mark-IV, Toon (1991); C01 – Curdt et al. (2001); F76 – Feldman et al. (1976); F91 – Feldman & Doschek (1991); G09 – García-Hernández et al. (2009); H58 – Herzberg (1958); J66 – Johansson (1966); K63 – Keenan & Greenstein (1963); K66 – Kaufman & Ward (1966); K98 – Klein et al. (1998); L95 – Liu et al. (1995); M81 – Mazzoni et al. (1981); P05 – Parenti et al. (2005); R27 – Ryde (1927); S47 – Shenstone (1947); S86 – Sandlin et al. (1986); SiC* – Newly observed lines from the SiC FT spectrum; TW – Either predicted with a better accuracy than that of Johansson (1966) in his Table 3 or newly calculated between energy level optimized in this work; W07 – Wallace & Hinkle (2007), followed by the different origin of FT spectra from the NSO archive as #1 – 840210R0.001, #2 – 810812R0.002, #6 – 880413R0.006, #13 – 850905R0.013. See text in Section 2.1 for more details. An extra '*' denotes either a newly measured line or the previous identification revised in this work. See Table 2 for revised energy levels.; W63 – Wilkinson & Andrew (1963); W96 – Wallace et al. (1996); Y91 – Yamamoto & Saito (1991)

[k] Number of sources, if more than one, used to obtain an averaged intensity.

[l] Comments: C – uncertainty of the line is the differences between the fitted wavelength and the line's center of gravity; D, E – the given uncertainty was doubled or tripled, respectively, compared to the original value in the quoted source; F – FT measurement; G – grating measurement; P – predicted line; S – single line that solely determines the upper energy level; T – intensity much greater than expected; U – intensity varies by an order of magnitude or more in different observations; V – intensity could not be reduced to the common-scale; W – intensity is much weaker than expected; X – the line was excluded from the level optimization; Y – blending reported in the original quoted work is removed in this work.

(The table is available in its entirety in a machine-readable form in the online journal. A few columns are omitted in this condensed table, but their footnotes ([c],[d],[f],[k]) are retained for guidance regarding their form and content.)

### *2.1. Measurements of Wallace and Hinkle*

As mentioned briefly above, reported wavenumbers of Wallace & Hinkle (2007) originate from three (out of four) spectrograms archived at the Virtual Solar Observatory (VSO) repository of the Kitt Peak National Observatory (KPNO), Tucson, AZ, where a 1 m (f/55 IR-visible-UV) FT spectrometer facility works in conjunction either with the McMath telescope's main beam or with laboratory sources (Brault 1978). We obtained all those spectrograms from VSO (Hill et al. 2004), achieved as 1981/08/12R0.002, 1984/02/10R0.001, 1985/09/05R0.013 and 1988/04/13R0.006 (hereafter called 81R02, 84R01, 85R13 and 88R06, respectively), and re-measured them as it was necessitated by the lack of detailed data on wavenumber measurement uncertainties in Wallace & Hinkle (2007). Those authors gave estimates of the systematic correction factors (with their uncertainty) for FTS-measured wavenumbers, and their table contains corrected wavenumbers and relative intensities. However, to determine the statistical uncertainties, we also need to know the line width and signal-to-noise ratio (*S/N*) for each line, which was not included by Wallace and Hinkle. Their paper also lacks any indication of the possible blending or distortions of observed line profiles. We tried to use the limited general information about the line widths given by Wallace and Hinkle to derive estimates of statistical uncertainties, but all these attempts led to large inconsistency between the observed and Ritz wavenumbers for many lines. Thus, we repeated the reduction of all available FTS recordings, including calibration, determination of the *S/N* ratio, and careful characterization of each line profile.

Among the four spectrograms, 85R13 contains the largest number of carbon lines (more than 270). It was taken by P.F. Bernath, using a microwave ($\mu$) discharge in a helium/methane mixture (2.75/0.03 Torr or 367/4 Pa) with an addition of phosphorous, at a resolution of 0.02 cm$^{-1}$ in the (1630–9860) cm$^{-1}$ region. The high-resolution (~0.01 cm$^{-1}$) spectrogram 84R01, taken by J.W. Brault with a carbon/Ne/CO (3.2/0.15 Torr or 430/20 Pa) hollow cathode lamp, covers nearly the same spectral region and has a significant number of carbon lines, but their signal strength is low compared to the $\mu$-discharge spectrum. Further down to the red end of the spectrum, the region (8995–16 010) cm$^{-1}$ is covered by the spectrogram 88R06 recorded by P.F. Bernath with a helium/allene (1.4/0.1 Torr or 190/13 Pa) $\mu$-discharge at a resolution of 0.02 cm$^{-1}$. It contains about 20 lines of C I. The last spectrogram, 81R02, was acquired by J.W. Allen with an electrodeless discharge in a Ne/methane (1.1/0.013 Torr or 150/1.7 Pa) mixture at a 0.03 cm$^{-1}$ resolution. It covers the optical region (15 477–27 000 cm$^{-1}$), but it has only a few C I lines.

**Table 2**

Observed Energy Levels of C I

| Configuration [a] | Term [a] | $J$ [a] | $E_{obs}$ ($cm^{-1}$) | Unc.($D_1$) [b] ($cm^{-1}$) | Unc. [b] ($cm^{-1}$) | Leading Percentages [c] | | | | | $\Delta E$[d] ($cm^{-1}$) | $LS$ [a] | No.L.[e] | Comment [f] |
|---|---|---|---|---|---|---|---|---|---|---|---|---|---|---|
| $2s^22p^2$ | $^3P$ | 0 | 0.0000000 | 0.0000006 | 0.0013 | 98 | | | | | −12 | | 51 | |
| $2s^22p^2$ | $^3P$ | 1 | 16.4167130 | 0.0000005 | 0.0013 | 98 | | | | | −9 | | 120 | |
| $2s^22p^2$ | $^3P$ | 2 | 43.4134567 | 0.0000007 | 0.0013 | 98 | | | | | −9 | | 141 | |
| $2s^22p^2$ | $^1D$ | 2 | 10192.657 | | 0.003 | 98 | | | | | 11 | | 76 | |
| $2s^22p^2$ | $^1S$ | 0 | 21648.030 | | 0.003 | 94 | 4 | $2p^4\ ^1S$ | | | 22 | | 49 | |
| $2s2p^3$ | $^5S°$ | 2 | 33735.121 | | 0.018 | 100 | | | | | 3 | | 2 | |
| $2s^22p3s$ | $^3P°$ | 0 | 60333.4476 | | 0.0005 | 96 | | | | | −66 | | 12 | |
| $2s^22p3s$ | $^3P°$ | 1 | 60352.6584 | | 0.0003 | 96 | | | | | −68 | | 34 | |
| … | … | … | … | … | … | … | … | … | … | … | … | … | … | … |
| **$2s^22p3p$** | **$^3P$** | **2** | **71385.40992** | **0.00006** | **0.00000** | **98** | | | | | **23** | | **43** | **B** |
| … | … | … | … | … | … | … | … | … | … | … | … | … | … | … |
| $2s^22p(^2P°_{1/2})5g$ | $^2[7/2]°$ | 3 | 86426.7910 | | 0.0006 | 99 | | | | | 1 | | 4 | |
| $2s^22p(^2P°_{1/2})5g$ | $^2[7/2]°$ | 4 | 86426.7917 | | 0.0006 | 99 | | | | | 1 | | 3 | |
| $2s^22p(^2P°_{1/2})5g$ | $^2[9/2]°$ | 5 | 86427.2556 | | 0.0003 | 99 | | | | | 1 | | 2 | R |
| $2s^22p(^2P°_{1/2})5g$ | $^2[9/2]°$ | 4 | 86427.25603 | 0.00020 | 0.0004 | 99 | | | | | 1 | | 2 | R |
| $2s^22p$5d | $^1$F° | 3 | 86449.208 | | 0.022 | 92 | 4 | $2p5d\ ^3F°$ | | | −3 | | 4 | |
| … | … | … | … | … | … | … | … | … | … | … | … | … | … | … |
| $2s^22p(^2P°_{3/2})$29d | $^2[7/2]°$ | 3 | 90753.8 | | 0.3 | | | | | | | $^1F°$ | 1 | |
| **C II ($2s^22p\ ^2P°_{1/2}$)** | **limit** | | **90820.348** | | **0.009** | | | | | | | | | |
| **C II ($2s^22p\ ^2P°_{3/2}$)** | **limit** | | **90883.743** | | **0.009** | | | | | | | | | |
| $2s2p^2(^4P)3s$ | $^5P$ | 1 | 103541.69 | | 0.24 | 100 | | | | | −1 | | 1 | |
| $2s2p^2(^4P)3s$ | $^5P$ | 2 | 103562.31 | | 0.10 | 100 | | | | | 1 | | 1 | |
| $2s2p^2(^4P)3s$ | $^5P$ | 3 | 103587.18 | | 0.15 | 100 | | | | | −2 | | 1 | |
| $2s2p^3$ | $^1D°$ | 2 | 103762 | | 19 | 95 | | | | | | | | L |
| … | … | … | … | … | … | … | … | … | … | … | … | … | … | … |
| $2s2p^2(^4P)14p$ | $^3D°$ | 3 | 133256.0 | | 1.8 | | | | | | | | 1 | |

**Notes:**

[a] The level designations are in either *LS* or *JK* (pair) coupling scheme. *JK*-designation is given for all previously known regular series 2*pnd* ($n \geq 7$) and 2*pnℓ* ($n \geq 8$, $\ell = s, p$) and their old *LS* designation with a leading percentage is given in column "*LS*".

[b] The quantity given in column "Unc." is the uncertainty of separation from the "base" level $2s^22p3p\ ^3P_2$ at 71 385.40992 $cm^{-1}$ (see text). The quantity in column "Unc. ($D_1$)" approximately corresponds to the minimum uncertainty of separation from other levels (for a strict definition, see Kramida (2011); if blank, it is the same as "Unc."). To roughly estimate an uncertainty of any energy interval (except those within the ground term), the values in column "Unc," should be combined in quadrature (see text in Section 3).

[c] The first leading percentage refers to the configuration and term given in the first two columns. The 2nd and 3rd percentages refer to the configuration and term subsequent to them. Percentages are blank for levels that were not included in the calculations.

[d] Differences between $E_{obs}$ and those calculated in the parametric least-squares fitting (LSF). Blank for unobserved levels or those excluded from the LSF or not included in the calculations.

[e] Number of observed lines determining the level in the level optimization procedure. Blank for unobserved levels.

[f] B – the base level for presentation of uncertainties; E – the energy level is extrapolated from the known quantum defects; L – the level value obtained in the parametric LSF calculation with Cowan's codes (see text); R – the value of $E_{obs}$ of previously unresolved fine-structure components is resolved in this work; T – the level position is tentative, based on a single line with an uncertain identification (see Table 1).

(This table is available in its entirety in a machine-readable form in the online journal. A portion is shown here for guidance regarding its form and content.)

The wavenumbers, $S/N$ ratios, line widths and intensities were determined from the FTS spectrograms with the help of the DEMCOP program (Brault & Abrams 1989) implemented for the X-windows graphical environment of Unix-based operating systems in the code XGREMLIN (Nave et al. 2015), which can find a least-squares fit to the Voigt profile of each line. In FT spectra, the scale of measured wavenumbers ($\sigma_{\mathrm{meas}}$) must be corrected. Although a well-controlled He-Ne sampling laser produces a fairly linear wavenumber scale, there remains some degree of imperfection in the alignment of optical beams from the laser and the light source, along with effects due to the finite entrance aperture size. In this regard, a multiplicative correction factor is derived from a set of accurately known reference wavenumbers ($\sigma_{\mathrm{ref}}$), either of buffer-gas atomic lines or of molecular features present in the spectrum. The corrected wavenumber ($\sigma_{\mathrm{cor}}$) can be expressed as

$$\sigma_{\mathrm{cor}} = (1 + k_{\mathrm{eff}})\sigma_{\mathrm{meas}}\ , \qquad (1)$$

where $k_{\mathrm{eff}}$ is a weighted mean of individual correction factors from each reference line, $k_i = \frac{\sigma_{\mathrm{ref},i}}{\sigma_{\mathrm{meas},i}} - 1$, with weight $w_i$ equal to inverse square of the total uncertainty $\delta k_i$. The latter is a combination in quadrature of the relative statistical uncertainty in $\sigma_{\mathrm{meas}}$ and the relative uncertainty in $\sigma_{\mathrm{ref}}$. The uncertainty in $k_{\mathrm{eff}}$ is estimated by an expression

$$k_{\mathrm{eff}} = \left[\sum_i \{w_i + w_i^2 (k_{\mathrm{eff}} - k_i)^2\}\right]^{½} \Big/ \sum_i w_i \qquad (2)$$

given by Radziemski & Andrew (1965). We note that Eq. (2) is similar to the equation defining the uncertainty $D_1$ in the level optimization code LOPT (Kramida 2011) and is an empirical extension of the standard statistical expression for the uncertainty of a weighted mean, $\delta = (\sum_i w_i)^{-½}$. This extension empirically accounts for various irregular systematic effects, such as line blending, which are often present in spectroscopic measurements.

The global systematic uncertainty in wavenumbers is determined as $\delta\sigma_{sys} = \sigma \times \delta k_{eff}$. The statistical uncertainty can be obtained from the formula

$$\delta\sigma_{stat} = \frac{k_g W}{(S/N)\sqrt{N_W}}, \tag{3}$$

where $k_g$ is a scaling factor depending on the choice of the fitting function (close to 1.0), $W$ is the full linewidth at half maximum (FWHM), $S/N$ is the signal-to-noise ratio, and $N_W$ is the number of statistically independent data points in FWHM, defined as ratio of the measured FWHM to the instrumental resolution. The final total uncertainty of a measured wavenumber is the sum in quadrature of the statistical uncertainty and the global systematic uncertainty (Redman et al. 2014).

The infrared spectrogram 85R13 is under-resolved (i.e., $N_w < 4$), contains many atomic and molecular lines (mainly due to hydrogen, helium, oxygen, phosphorous, argon, and carbon monoxide) and additional noisy features of instrumental origin. The measurements were made after the background subtraction and interpolation of the spectrum, which doubled the number of data points in it. The calibration was made with recommended infrared standards of the 1–0 band of carbon monoxide (Maki & Wells 1992), and the systemic correction factor (obtained using 12 well-resolved reference lines) is $k_{eff} = 5.60(18)\times10^{-7}$, which is about the same as reported by Wallace & Hinkle (2007) ($5.8\times10^{-7}$). However, their systematic uncertainty was a factor of 3/2 greater than ours, as they used less accurate wavenumbers from Guelachvili & Narahari Rao (1986).

For the supplementary IR spectrogram 84R01, we obtained $k_{eff} = 5.5(3)\times10^{-7}$ using 23 low-excitation ($3p \rightarrow 4s$) reference lines of Ne I (Saloman & Sansonetti 2004). For both this spectrogram and 85R13 discussed above, we take $k_g = 1.5$, since the number of data points in the observed linewidth was small ($N_w \approx 1.5$).

Five low-excitation ($4s \rightarrow 4p$) Ar I lines (Sansonetti 2007) served as internal standards to calibrate the 88R06 spectrum in the (8995–16 010) $cm^{-1}$ region, resulting in $k_{eff} = -3(6)\times10^{-8}$. Even though Wallace & Hinkle (2007) used Ar I wavenumbers from Whaling et al. (2002) affected by an error in the systematic correction found later by Sansonetti (2007), our value essentially agrees with theirs, $+5(9)\times10^{-8}$.

The spectrogram 81R02 covering the region (15 477–27 000) $cm^{-1}$ was calibrated with seven strong $4s \rightarrow 5p$ lines of Ar I (Sansonetti 2007). A value $k_{eff} = -5.7(3)\times10^{-7}$ was determined in agreement with that of Wallace & Hinkle (2007), $-5.3(3)\times10^{-7}$. In addition to Ar I, this spectrum contains many strong Ne I lines. However, most of them are from higher excited levels (4*d*, 5*d*, 6*s*, 7*s*, 8*s*) and produce a different value of $k_{eff}$ (30 % smaller). This may be explained by a different spatial origin of these lines, and/or by isotope shifts (which are larger in Ne than in Ar), and/or by possible Stark shifts, which are stronger for higher excited levels. Therefore, these high-excitation Ne lines were not included in the calibration.

The corrected wavenumbers with their assessed uncertainties and assignments were then taken into a preliminary level optimization process. As mentioned above, since the spectrum 85R13 is under-resolved, the lines are affected by ringing or side lobes of strong nearby features, blending by known or unidentified lines, and asymmetries in the line shape. Those affected lines were double-checked at various stages of the level optimization process.

### *2.2. Observations of Johansson & Litzén and Johansson*

The measurements of Johansson & Litzén (1965) and Johansson (1966), made with a Fabry-Perot interferometer and a grating spectrograph, respectively, are an extensive source of atomic data for C I. The FT spectrogram 85R13 fully superseded the IR measurements of Johansson & Litzén (1965), whose estimate of uncertainty was better than 0.02 $cm^{-1}$. We re-assessed the uncertainty with the aid of Ritz wavenumbers from the FT spectra. It turned out that most of the wavenumbers reported by Johansson and Litzén agree with the Ritz ones with a standard deviation (hereafter, uncertainty) of 0.01 $cm^{-1}$, except for unresolved features.

The other set of measurements of Johansson (1966) was made with two different grating spectrographs. For wavelengths $\lambda < 9600$ Å, he used a 21 ft. concave grating spectrograph with a reciprocal linear dispersion (hereafter, dispersion) 5 Å/mm in the first order of diffraction. Wavelengths below 4700 Å were measured in the second order of this instrument. For the wavelengths in the range of (10 000–11 600) Å, a plane grating spectrograph was used in the second order, where it had a dispersion of ~2 Å/mm. Johansson reported wavelengths ~380 C I lines in the (2478–11 331) Å range, most of which were given with three digits after the decimal point. We selected 104 uniquely classified and well-resolved lines having Ritz wavelength uncertainties in the range of (0.000 06–0.0018) Å to examine the wavelengths reported by Johansson in the three spectral regions described above. The estimated uncertainties are 0.005 Å, 0.014 Å and 0.009 Å for $\lambda < 4700$ Å, (4700–9600) Å, and $\lambda > 9600$ Å, respectively, and no significant systematic errors were found. These uncertainties are about a factor of two smaller than Johansson's estimates.

*2.3. Observations of Transitions within the Ground Configuration*

Forbidden transitions (M1 & E2) between the levels of the ground configuration $2s^2 2p^2$ are distributed in three regions, far infrared (FIR), NIR, and optical, as can be easily inferred from Figure 1. As mentioned above, some of these NIR and/or optical transitions were observed in several astrophysical objects (Boyce 1936; Lambert & Swings 1967; Swensson 1967; Liu et al. 1995). The implementation of FIR-laser magnetic resonance (FIR-LMR) technique (Saykally & Evenson 1980) brought up the first precise laboratory values for $^3P_0$–$^3P_1$ and $^3P_1$–$^3P_2$ transitions at ~492 and ~809 GHz, respectively for both $^{12}$C and $^{13}$C isotopes. We accepted the most precise and accurate measurements available now for $^3P_0$–$^3P_1$ (Yamamoto & Saito 1991) and $^3P_1$–$^3P_2$ (Klein et al. 1998) transitions. The reported uncertainties of these measurements were given on the level of two or three standard deviations, and we reduced them to the 1-sigma level to provide a uniform representation of uncertainties in all data.

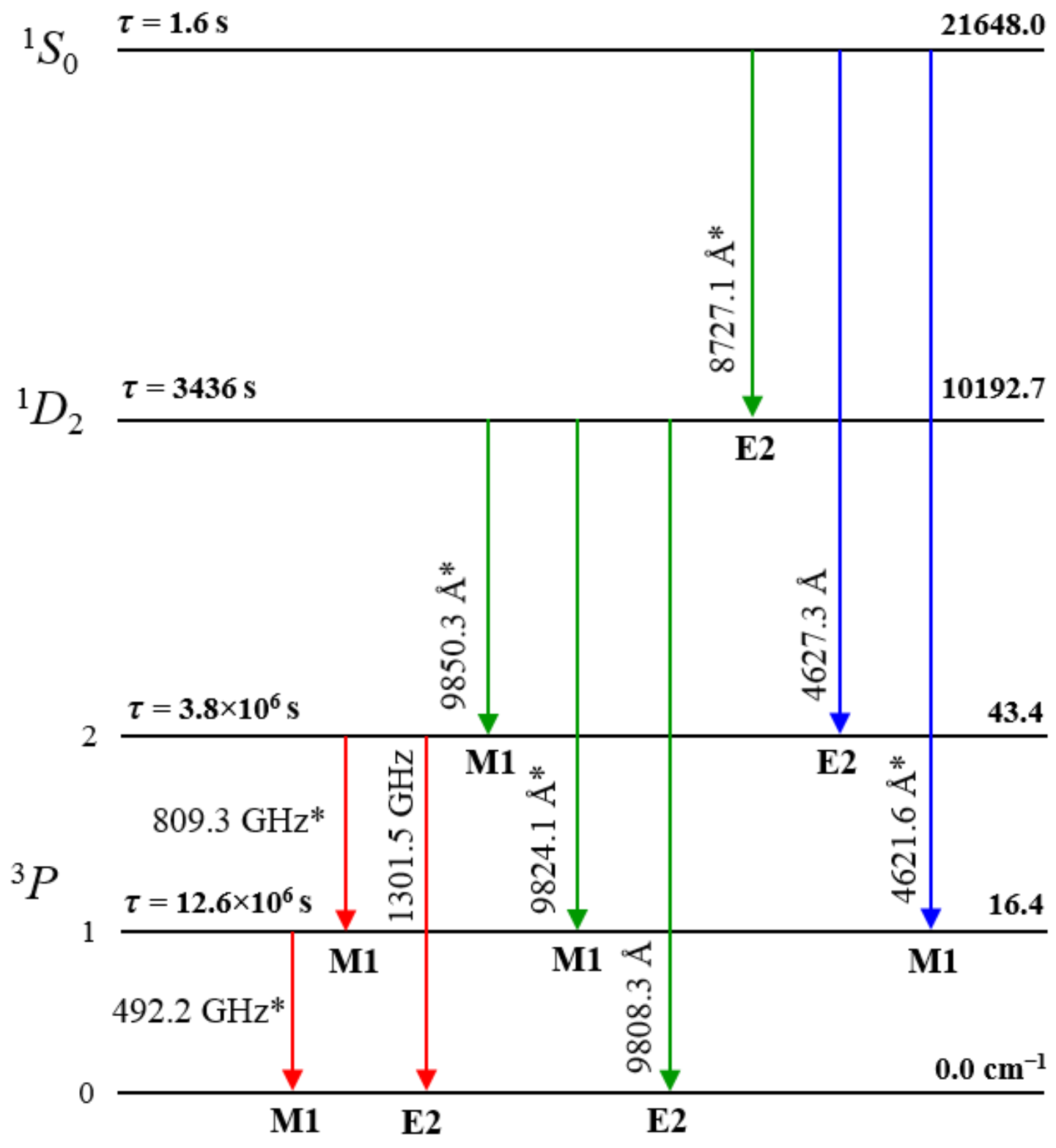

**Figure 1** (*Color online*). Energy level diagram and forbidden transitions within the levels of the $2s^22p^2$ configuration in C I. Transitions observed in laboratory and/or astrophysical sources are marked with an asterisk (see Table 3). The lifetime ($\tau$) is derived from *A*-values compiled by Wiese & Fuhr (2007). The values on the right are level energies in $cm^{-1}$.

One of the accurately measured transitions is $(2s^22p^2)^3P_1$–$^1S_0$ at 4621.5695(7) Å. This value is from our measurement of the 81R02 FT spectrogram described in section 2.1. This spectrogram was acquired with the lowest buffer-gas pressure among all other FT spectra. To validate an observation of a forbidden transition in such a laboratory spectrum, many considerations were taken into account. From the analysis of observed C I line intensities, it was found that the effective excitation temperature ($T_{eff}^{exc}$) in this spectrum was 0.28 eV, which is the lowest among the spectra from the same laboratory. The low lamp pressure and temperature imply that electron density was minimal, which favors observation of this forbidden transition. In this regard, we note that it requires about twice more energy to ionize the neon atoms than carbon. $T_{eff}^{exc}$ derived from lines of Ne I was 0.20 eV. The validity of both derived source temperatures was verified by comparing the wavelength-dependent registration-response functions derived from the C I and Ne I spectra, which agreed with each other. Under such low temperatures, most of the free electrons in the plasma originate from singly-ionized carbon atoms. Under LTE conditions, the fraction of $C^+$ in the discharge would be about 1 % (see the Saha-Boltzmann plot for a C+N mixture in the NIST ASD (Kramida et al. 2016), which should result in electron densities $n_e < 10^{14}$ $cm^{-3}$. This estimate is supported by a rough determination of $n_e$ from the measured Stark width $w_s$ of the hydrogen $H_\beta$ line at $\lambda = 4861.323$ Å (hydrogen was present as a small impurity in the spectra we analyzed). From the measured value of $w_s \approx 0.06$ Å we derived the value of $n_e$ of the order of $10^{13}$ cm$^{-3}$. Plausibility of an observation of forbidden lines within the ground configuration of C I under these conditions is further confirmed by the work of Eriksson (1965) who observed a similar forbidden transition within the ground configuration of O I with an electrodeless discharge lamp that had a similarly low pressure.

We searched for the two transitions from the same upper level and found a faint asymmetric line (with *S*/*N* just above 3) at 4627.3439(22) Å corresponding to the $^3P_2$–$^1S_0$ transition, which we adopted as a questionable identification, since the predicted intensity of this transition is two orders of magnitude lower than observed. No other forbidden lines were observed in the visible range, except the line at 8727.126(8) Å ($^1D_2$–$^1S_0$) reported by Wallace et al. (1996) from solar observations. Among the NIR transitions, the $^3P_1$–$^1D_2$ and $^3P_2$–$^1D_2$ lines were observed in several astrophysical objects (Swensson 1967; Liu et al. 1995), and their averaged observed wavelengths are 9824.31(19) and 9850.34(9) Å, respectively. All these observations are summarized in Table 3 with their transition rates and accurate Ritz values.

**Table 3**

Observed Astrophysically Important Forbidden Transitions of C I ($^{12}$C/$^{13}$C) from Laboratory and Astrophysical Sources

| Transition | Species | Laboratory [a] | Astronomical [b] | Ritz [c] | $A$ (s$^{-1}$) [d] | | Ref. [e] |
|---|---|---|---|---|---|---|---|
| | $^{12}$C | 492160.651(18) | | | | | Y91 |
| $^3P_0$–$^3P_1$ | $^{13}$C | 492162.900(35) | | 492162.889(18) | 7.93e−08 | A | Y91 |
| | C | 492160.675(19) | 492160.7(1) | 492160.675(18) | | | Y91, F89 |
| | $^{12}$C | 809341.970(17) | | | | | K98 |
| $^3P_1$–$^3P_2$ | $^{13}$C | 809346.1(1) | | 809346.103(44) | 2.65e−07 | A | K98 |
| | C | 809342.01(2) | 809342.3(4) | 809342.014(17) | | | K98, K98a |
| | $^{12}$C | | | 1301502.621(25) | | | |
| $^3P_0$–$^3P_2$ | $^{13}$C | - | - | 1301508.990(48) | 1.72e−14 | B+ | - |
| | C | | | 1301502.689(25) | | | |
| $^3P_2$–$^1D_2$ | C | - | 9850.34(9) | 9850.250(3) | 2.2e−04 | C+ | L95 |
| $^3P_1$–$^1D_2$ | C | - | 9824.31(19) | 9824.118(3) | 7.3e−05 | C+ | |
| $^3P_0$–$^1D_2$ | C | - | - | 9808.295(3) | 5.9e−08 | C | - |
| $^1D_2$–$^1S_0$ | C | - | 8727.126(8) | 8727.131(3) | 6.0e−01 | B | W96 |
| $^3P_2$–$^1S_0$ | C | 4627.3438(22)[f] | - | 4627.3444(6) | 2.2e−05 | C | TW |
| $^3P_1$–$^1S_0$ | C | 4621.5695(7) | 4261.4 | 4621.5693(6) | 2.3e−03 | C+ | TW, B36 |

[a] For all transitions, except the first three, observed and Ritz wavelength (in standard air) are given in Å. For the first three transitions, observed frequencies are given in MHz. Multiple values given in consecutive rows are for $^{12}$C, $^{13}$C and natural C; a single value represents natural C samples; the value in parentheses is the 1-$\sigma$ uncertainty in the last significant digit.
[b] The best available astronomical data, the uncertainties are from the original measurements.
[c] The Ritz values from the level optimization made in this work (see Sections 3 and 4).
[d] The transition rate (*A*-value) with the stated accuracy in the next column are from Wiese & Fuhr (2007). The accuracy code is described in Table 10.
[e] Reference code for the values in column 2 and 3, respectively: B36 – Boyce (1936); F89 – Frerking et al. (1989); K98 – Klein et al. (1998); K98a – Keene et al. (1998); L95 – Liu et al. (1995); TW – This work, from the 81R02 FT spectrum; Y91 – Yamamoto & Saito (1991); W96 – Wallace et al. (1996).
[f] The identification is questionable.

## *2.4. Vacuum ultraviolet FTS measurements*

Griesmann & Kling (2000) made accurate interferometric measurements of lines of carbon and other elements in the VUV range. They kindly permitted us to use their unpublished data obtained during that work, in which we found a few C I lines. Their measurements were made with the FT700 VUV FT spectrometer at NIST (Griesmann et al. 1999) using a Penning discharge with silicon carbide (SiC) cathodes and Ne as carrier gas. Unpublished data of Griesmann and Kling included recordings of two interferograms containing features of neutral carbon. One of them had a resolution $R = 0.05$ cm$^{-1}$ and the calibration factor determined by Griesmann and Kling was $7.16(12)\times10^{-7}$. For the second interferogram with $R = 0.25$ cm$^{-1}$, covering the range (52 000–78 000) cm$^{-1}$, we derived the calibration factor of $1.95(6)\times10^{-6}$ from Si II lines. Statistical uncertainties of these measurements are in the range (0.0006–0.017) cm$^{-1}$, but there is a large systematic component in the total uncertainty. The second spectrum, most useful for our purpose, contains eleven accurately measured lines of the $2s^22p^2\ ^3P$ – ($2p3s\ ^3P°+2s2p^3\ ^3D°$) transitions in addition to three lines measured from the first recording. The earlier best measurements for these VUV lines were made with grating spectroscopy by Kaufman & Ward (1966). By comparing their reported wavelength with our FTS measurements and with Ritz wavelength, we found that measurements of Kaufman and Ward had a small systematic shift of about −0.0006 Å, which we removed from their reported wavelengths. After that, the remaining statistical uncertainty of a few lines adopted from Kaufman and Ward is estimated to be 0.0008 Å.

*2.5. Observations of Chang and Geller*

The C I lines reported by Chang & Geller (1998) originated from four sets of FTS measurements of solar spectra. Some of them were space-based, some balloon-borne, and some ground-based. The largest number of lines, 148, are from the space-based ATMOS mission and cover the (600–4800) $cm^{-1}$ region (Farmer & Norton 1989). About 40 C I lines are from the balloon-borne MARK IV observation (Toon 1991). Both these spectra are reasonably free from telluric contamination. The other two observations are from the McMath telescope in conjunction with the 1 m FTS of KPNO, in the regions of (1850–9000) $cm^{-1}$ (Livingston & Wallace 1991, hereafter NOAO1) and (8900–13 600) $cm^{-1}$ (Wallace et al. 1993, hereafter NOAO2). Most of these measurements were superseded by laboratory FTS measurements. However, extraordinary conditions could exist in the solar atmosphere, making it possible to observe some transitions that could not be observed in laboratory light sources. Indeed, we found such transitions in these solar spectra.

None of the original solar line lists mentioned above explicitly reported the measurement uncertainties. However, notes in Chang & Geller (1998) on possible blending and other factors affecting the measurements were helpful in our assessment of uncertainties, which was mainly based on comparisons with Ritz wavenumbers derived from laboratory spectra. The statistical uncertainty was estimated using equation (3), under the assumption that the weakest observed lines had $S/N = 1$, and that the line widths were largely due to Doppler broadening. The latter was estimated using the photospheric temperature $T_{phot} = 5500$ K, which is consistent with Boltzmann plots used in the intensity reduction (see Section 5). Many wavelengths were still deviating too much from the Ritz values, which forced us to increase the statistical uncertainty of many perturbed lines (blended, appearing on shoulders of stronger lines, or contaminated by telluric features) to half of their width irrespective of their intensity. The unweighted systematic correction factors and their uncertainties were estimated using accurately known C I Ritz wavenumbers. Corrections to the originally reported calibration were found to be much smaller than statistical uncertainties; hence all originally reported wavenumbers were retained in this work. For the ATMOS spectrum, we obtained a global systematic correction $k_{\mathrm{eff}} = 7(6)\times10^{-7}$ to wavenumbers listed by Chang and Geller from 19 reference lines. Statistical uncertainties of lines from this spectrum are in the ranges (0.006–0.015) $\mathrm{cm}^{-1}$ and (0.006–0.022) $\mathrm{cm}^{-1}$ for unperturbed and perturbed lines, respectively. The MARK IV spectrum was compared with 13 Ritz wavenumbers accurate to better than 0.008 $\mathrm{cm}^{-1}$; $k_{\mathrm{eff}} = 6.6(19)\times10^{-7}$ was derived. For this spectrum, statistical uncertainties were in the range (0.007–0.026) $\mathrm{cm}^{-1}$. The ground-based observations are more prone to atmospheric contamination. The correction factors and statistical uncertainties were found to be $k_{\mathrm{eff}} = -5.3(20)\times10^{-7}$, $\sigma_{\mathrm{stat}}$ = (0.0004–0.040) $\mathrm{cm}^{-1}$, and $k_{\mathrm{eff}} = +5.1(25)\times10^{-7}$, $\sigma_{\mathrm{stat}}$ = (0.0003–0.062) $\mathrm{cm}^{-1}$ for the NOAO1 and NOAO2, respectively. After combining both systematic and statistical uncertainties in quadrature, 74 lines from these four spectra were added to the C I linelist.

It is important to mention here that the ATMOS measurements were followed by similar solar measurements with the Atmospheric Chemistry Experiment (ACE) FTS onboard the Canadian SCISAT−1 satellite (Hase et al. 2010). The measurement accuracy of ACE-FTS was inferior to ATMOS because of the twice lower resolution. Nonetheless, the accuracy of observed line intensities was much improved by co-adding multiple recordings. Thus, the ACE-FTS data were used for intensity reduction.

*2.6. Observations of VUV transition arrays*

VUV transition arrays of C I were observed in many laboratory sources (Fowler & Selwyn 1928; Paschen & Kruger 1930; Edlén 1933b; Wilkinson 1955; Herzberg 1958; Wilkinson & Andrew 1963; Kaufman & Ward 1966; Junkes et al. 1965; Mazzoni et al. 1981), as well as in several solar prominences (Sandlin et al. 1986; Feldman & Doschek 1991; Curdt et al. 2001; Parenti et al. 2005; Feldman et al. 1976). These measurements were compared with Ritz wavelengths (accurate to better than 0.0003 Å) generated using long-wavelength FTS measurements in combination with several VUV lines precisely measured with an FTS (see Sections 2.1–2.5). Evaluated uncertainties of individual measurements with a brief description of observations are tabulated in Table 4. No significant systematic shifts were found in these measurements.

**Table 4**

List of Vacuum Ultraviolet Measurements in C I with Their Evaluated Uncertainties

| **Source** [a] | **No. of lines** | **Range (Å)** | **Unc.** [b] **(mÅ)** | **Description** [c] | $\delta\lambda/\delta x$[d] **(Å/mm)** |
|---|---|---|---|---|---|
| F28 | 15 | 1260–1752 | 60 | $CO_2$-He; G; Ph | 10 |
| P30 | 94<br>21 | 1112–1400<br>1400–2950 | 50<br>50 | He-C HC; 1-m G; Ph | 8.5<br>17.5 |
| E33b | 15 | 945–1659 | 20 | Vacuum spark; 1-m G; Ph | 8.3 |
| B36 | 100 | 945–1994 | 30 | Gas discharge; 2-m G; Ph | 4.3 |
| S47 | 9<br>3 | 1430–1660<br>2582–2968 | 10<br>2 | Carbon arc; 21-ft G; Ph | 1.3 |
| W55 | 43 | 1158–1931 | 5 | HC; 21-ft G; Ph | 1.3 |
| H58 | 6 | 1328–1330 | 0.7 | HC; 3-m G; Ph | 0.6{5} |
| W63 | 14 | 1560–2478 | 3 | HC; 21-ft G; Ph | 0.4{3} |
| J65 [e] | 180 | 1116–2252 | – | HC; 1-m G; Ph | 8.1 |
| K66 [f] | 18 | 1459–1931 | 0.8 | HC; 10.7-m G; Ph | 0.78 |
| F76 | 120 | 1100–1930 | 10; 4 | Solar flare; 2-m G; Ph | 4.2{2} |
| C81 | 33 | 750–870 | 10 | AFP; 2-m G; Ph | 4.2 |
| M81 | 150 | 1100–1161<br><1200 | 10; 5<br>7 | | |
| S86 | 200 | 1188–1660 | 10; 3 | SP; 2-m G; Ph | 4.2{2} |
| F91 | 45 | 1114–1160 | 10 | | |
| C01 | 216 | 945–1610 | 10 | SP; 3.2-m G; MCP | 0.87 |
| P05 | 100 | 945–1260 | 20–40 [g] | | |
| L05 | 3 | ~945 | 0.06 [h] | $C_2H_2$/He; tunable XUV laser | – |

[a] Source reference code: B36 – Boyce & Robinson (1936); C81 – Cantù et al. (1981); C01 – Curdt et al. (2001); E33b – Edlén (1933b); F28 – Fowler & Selwyn (1928); F76 – Feldman et al. (1976); F91 – Feldman & Doschek (1991); H58 – Herzberg (1958); J65 – Junkes et al. (1965); K66 – Kaufman & Ward (1966); L05 – Labazan et al. (2005); M81 – Mazzoni et al. (1981); P30 – Paschen & Kruger (1930); P05 – Parenti et al. (2005); S47 – Shenstone (1947); S86 – Sandlin et al. (1986); W55 – Wilkinson (1955); W63 – Wilkinson & Andrew (1963).
[b] Uncertainty evaluated by comparing observed and Ritz wavelengths of unperturbed lines (see text). When two figures separated by semicolon are given, they correspond to wavelengths reported with 2 and 3 decimals after the point, respectively. For perturbed lines (e.g., blended, doubly classified, or shaded) the given uncertainty is doubled.
[c] A brief experimental description of the light source, instrument and detector used. The symbols HC, G, Ph, Q, AFP, SP, and MCP denote hollow cathode discharge, grating, photographic plate/film, quartz prism, absorption with flash pyrolysis, solar prominences, and microchannel plate, respectively.
[d] Inverse linear dispersion (in the first order of diffraction unless the order is specified in curly braces).
[e] Only intensities were measured, but not wavelengths.
[f] A systematic shift of –0.0006 Å was removed from the originally reported values (see Section 2.4).
[g] The given value is quoted from the original work. Our comparison shows an average uncertainty of 30 mÅ.
[h] Uncertainty could not be estimated in this work; the original value is quoted.

## *2.7. Observations of Other Transitions*

The spin-forbidden (intercombination) $2s^22p^2\ ^3P_{1,2} – 2s2p^3\ ^5S°_2$ transitions near 2966 Å were first measured by Shenstone (1947) with an uncertainty of 0.002 Å and later remeasured for the $^{12}$C and $^{13}$C isotopes by Bernheim & Kittrell (1980) in the 19$^{th}$ order of a grating spectrometer. Once the most accurate measured wavelengths were selected for each firmly identified transition, and their uncertainties were evaluated, we reoptimized the energy levels and searched for possible identification of previously unclassified lines in published line lists. We found many suspected C I lines in the list of lines observed by Keenan & Greenstein (1963) in a spectrum of a carbon-rich star R Coronae Borealis (R-CrB). About 240 C I lines were already identified by those authors based on laboratory observations (Johansson 1966), but a possible extension of the 2*p*3*s* → *np* and 2*p*3*p* → (*ns*+*nd*) series could be expected due to peculiarity of stellar conditions. These measurements were primarily divided into two sections, below $\lambda$ = 4900 Å and above that. The estimated uncertainties are 0.10 Å and 0.20 Å (for unperturbed and perturbed lines, respectively) for the short wavelengths and 0.16/0.28 Å, respectively, for the longer wavelengths. The quoted uncertainties are twice the standard deviations (SD) of observed wavelengths from the Ritz values, allowing for possible unreported blending in this spectrum, strongly contaminated with lines of many chemical elements and molecules.

### *2.8. Autoionizing States*

Data on autoionizing states of carbon are important for studies of radiation transport in stars. The ionization continuum of C I essentially starts from the first series limit, C II $2s^22p\ ^2P^o_{1/2}$ at 90 820.348 $cm^{-1}$ (see Section 7). Among the regular configurations involving excitation of one $2p$ electron, only high Rydberg states of $2s^22pn\ell$ converging to the second limit (C II $2s^22p\ ^2P^o_{3/2}$ at 90 883.743 $cm^{-1}$) are autoionizing. All other excited configurations, $2s2p^2n\ell$, $2p^3n\ell$, $2p^4$, and those involving excitation from the inner $1s$ shell lie above the ionization threshold, except $2s2p^3$, in which the $^5S^o$, $^3D^o$ and $^3P^o$ levels are bound, but $^3S^o$, $^1D^o$, and $^1P^o$ are above the ionization limit.

Not all states above the first limit are easily decaying by autoionization. For some of them, autoionization is strongly forbidden by selection rules (for example, quintet states), and some have competing rates for autoionization and radiative decay. The first observation of such radiative transitions was made for the decay of $2s2p^3\ ^3S^o_1$ to the ground $^3P$ term near 945 Å by Edlén (1933b). The wavelengths of these transitions are now known with an accuracy of 6 parts in $10^8$ (Labazan et al. 2005). In pure $LS$ coupling, the $2s2p^3\ ^3S^o_1$ state would be strictly forbidden to autoionize. In fact, autoionization is possible due to a minute admixture of the $^3P^o$ and $^1P^o$ character in its eigenvector composition. The total radiative decay rate of $3.41\times10^9$ $s^{-1}$ is comparable to the autoionization rate of $1.50\times10^9$ $s^{-1}$ (Wang et al. 2013). Autoionization of $2s2p^3\ ^3S^o_1$ is responsible for about 30 % of its total decay rate and significantly contributes to the observed widths of radiative transitions, ≈1 GHz (Labazan et al. 2005). The smallness of the autoionization rate of this state enables observation of other radiative transitions terminating on $2p3p\ ^3P$, which Johansson (1966) identified near 2903 Å in the spectrum of a high-current arc observed by Ryde (1927). The observations by Goly (1976), who measured radiative rates of these lines, further substantiated the above identification.

Transitions from the quintet term $2s2p^23s\ ^5P$ to $2s2p^3\ ^5S^o_2$ were found by Shenstone (1947) near 1432 Å, which helped him to identify two observed lines near 2966 Å as intercombination transitions between $2p^2\ ^3P_{1,2}$ and $2s2p^3\ ^5S°_2$, and hence to locate the latter level precisely. Edlén (1947) confirmed this identification by an isoelectronic comparison.

Thirty-three strong absorption features due to the $2s^22p^2\ ^3P$–$2s2p^2np\ ^3D°$ ($n$ = 3–14) transitions were observed by Cantù et al. (1981) in the photoionization spectrum between 750 Å and 870 Å. Of those, the $^3D°_1$ and $^3D°_3$ series were the most extensive. The $2s2p^2np\ ^3D°$ series converge to the C II $2s2p^2\ ^4P$ limits at about 43 003 cm$^{-1}$ above the ground state of C II, $2s^22p\ ^2P°_{1/2}$, while the C II $2s^22p\ ^2P°_{3/2}$ level has an excitation energy of only ~63 cm$^{-1}$. Therefore, the final state of autoionization decay of the C I $2s2p^2np\ ^3D°$ levels can be either of these two C II levels. Due to constraints on excitation conditions, most laboratory photoionization data on C I are for excitation from the ground $^3P$ term to continuum states. However, contributions of excitation from $^1D$ and $^1S$ (of $2s^22p^2$) to, for example, $2s2p^3\ ^1D^o$ and $^1P^o$ embedded in the continuum can be significant in astrophysical objects. Complexities in producing a high temperature vapor of neutral carbon explain scarcity of experimental photoabsorption studies, of which Hofmann & Weissler (1971) and Marrone & Wurster (1971) are rare exceptions. So far, no experimental data on the exact position of the strongly autoionizing $2s2p^3\ ^1D^o$ and $^1P^o$ states were obtained, either by optical or electron spectroscopy. The AEL (Moore 1970) gave their respective energy positions at 97 878 cm$^{-1}$ and 119 878 cm$^{-1}$. These values were derived by Edlén (1933b) from an isoelectronic extrapolation. We used the present parametric least-squares fits to pin down their positions, which turned out to be 103 762 cm$^{-1}$ and 115 209 cm$^{-1}$, respectively, for the $2s2p^3\ ^1D^o$ and $^1P^o$ states (with an estimated uncertainty of 19 $cm^{-1}$). Accordingly, the predicted wavelengths of their decay to $^1D$ and/or $^1S$ of the ground configuration are 1068.7 Å for $2s2p^3\ ^1D^o$ and 952.2/1068.8 Å for $^1P^o$. No observed features were found near these wavelengths in either Hofmann & Weissler (1971) or Marrone & Wurster (1971). In the first of these studies, the region (1048–1068) Å was masked by strong Ar I resonance lines. The second one reported photoionization cross-section only at four points of the continuum (982, 1005, 1020 and 1060) Å and hence was not helpful for the line search. A complementary electron-spectroscopy data source is a very low-energetic Auger spectrum taken by Lee & Edwards (1975) in the (2.23–8.74) eV region. The lower energy (2.23 eV) sets a threshold of 108 807 $cm^{-1}$ for the lowest detectable level. Hence, $^1D^o$ could not be observed. Nevertheless, 18 peaks reported in their spectrum produced by collisions of carbon with helium contain signatures of states with an open $2s$ subshell.

In spite of existence of many series converging to various limits, the first excited term in C II is near 43 003 $cm^{-1}$ (or 5.33 eV), which allows all autoionizing states of C I with energy less than 133 823.8 $cm^{-1}$ (16.59 eV) to decay only to the ionic ground term $2s^22p\ ^2P°$. Due to limited resolution of the measurements of Lee & Edwards (1975), we could tentatively identify only some of the features listed in their Table 1 using level energies and autoionizing rates calculated in our parametric LSF with Cowan's codes (Cowan 1981). The first peak at 2.23 eV can be assigned to the fast-autoionizing term $2s2p^2(^4P)3s\ ^3P$ whose energy from the LSF is at 108 100 $cm^{-1}$. A few of the original identifications of Lee and Edwards agree with the photoionization study carried out later by Cantù et al. (1981). In order to observe $2s2p^3\ ^1P^o$ at 14.28 eV (as predicted by our LSF), an ejected-electron energy of 3.02 eV should be observed, but it cannot be resolved from the very strong photoionization feature of $2s^22p^2(^4P)3p\ ^3D°$ at 2.96 eV. Thus, the position of the former term is yet uncertain.

## Table 5

Revision of Assignments of Autoionizing States of C I Reported by Lee & Edwards (1975)

| $E_p$ [a] (eV) | Label [b] | Assignment [c] | | | $E$ [d] (eV) | Comment [e] |
|---|---|---|---|---|---|---|
| 2.23 | a | $2s2p^2(^4P)3s\ ^3P$ | – | $2s^22p\ ^2P^\circ$ | 13.40 | R |
| 2.36 | m′ | $2s2p^2(^2D)3d$ | – | $2s2p^2\ ^4P$ | 19.02 | R |
| | | $2s2p^2(^2P)3d$ | – | $2s2p^2\ ^2D$ | | |
| 2.62 | b | $2s2p^2(^2P)3p\ ^5D^\circ$ | – | $2s^22p\ ^2P^\circ$ | 13.85 | N |
| | | $2s2p^2(^2D)4p$ | – | $2s2p^2\ ^4P$ | | |
| 2.72 | n′ | $2s2p^2(^2P)3p\ ^5P^o$ | – | $2s^22p\ ^2P^\circ$ | 13.92 | R |
| | | $2s2p^2(^2D)4p$ | – | $2s2p^2\ ^4P$ | | |
| 2.96 | c | $2s2p^2(^4P)3p\ ^3D^o$ | – | $2s^22p\ ^2P^\circ$ | 14.27# | A, C |
| | | $2s2p^3\ ^1P^o$ | | | 14.28 | |
| 3.11 | o′ | $2s2p^2(^4P)3p\ ^3P^o$ | – | $2s^22p\ ^2P^\circ$ | 14.43 | R |
| 3.29 | d | $2s2p^2(^4P)3p\ ^3S^o$ | – | $2s^22p\ ^2P^\circ$ | 14.42 | R |
| 3.72 | e | $2s2p^2(^4P)3d$ | – | $2s^22p\ ^2P^\circ$ | 15.03 | A |
| | | $2s2p^2(^4P)4s$ | | | 14.95 | |
| 4.16 | f | $2s2p^2(^4P)4p\ ^3D^o, ^3P^o$ | – | $2s^22p\ ^2P^\circ$ | 15.44# | C |
| 4.42 | g | $2s2p^2(^4P)4d$ | – | $2s^22p\ ^2P^\circ$ | 15.72 | |
| 4.60 | h | $2s2p^2(^4P)5p\ ^3D^o$ | – | $2s^22p\ ^2P^\circ$ | 15.90# | A, C |
| | | $2s2p^2(^4P)4f$ | | | 15.84 | |
| 4.74 | i | $2s2p^2(^4P)5d\ ^3D$ | – | $2s^22p\ ^2P^\circ$ | 16.05 | |
| 5.51 | j | $2s2p^2(^2D)3s\ ^3D$ | – | $2s^22p\ ^2P^\circ$ | 17.00 | |
| 6.68 | k | $2s2p^2(^2D)3p$ | – | $2s^22p\ ^2P^\circ$ | 18.00 | |
| 7.74 | m | $2s2p^2(^2D)3d$ | – | $2s^22p\ ^2P^\circ$ | 19.02 | |
| | | $2s2p^2(^2D)4p$ | | | 19.15 | |
| 8.06 | n | $2s2p^2(^2D)4p$ | – | $2s^22p\ ^2P^\circ$ | 19.25 | A |
| | | $2s2p^2(^2D)4s$ | | | | |
| 8.42 | o | $2s2p^2(^2D)4f$ | – | $2s^22p\ ^2P^\circ$ | 19.60 | R |
| | | $2s2p^2(^2D)5p$ | | | 19.70 | |
| 8.74 | q | $2s2p^2(^2S)3s\ ^1S$ | – | $2s^22p\ ^2P^\circ$ | 19.95 | R |
| | | $2s2p^2(^2D)4d$ | | | | |

[a] $E_p$ refers to observed energy of an Auger-electron peak with an uncertainty of 0.07 eV.

[b] Peak labels quoted from the original work (see Table I and figure 2 in Lee & Edwards 1975).

[c] The first refers to an excited state of C I decaying to a residual state of C II in the next column. A few peaks are multiply classified. If no terms are given, it means there are several contributing terms, mostly of triplet character. The excitation thresholds of the C II $2s^22p\ ^2P^\circ$, $2s2p^2\ ^4P$ and $2s2p^2\ ^2D$ terms are 11.26 eV, 16.60 eV, and 20.55 eV, respectively (relative to the ground level of neutral C).

[d] Mean energy of the autoionizing state indicated. Observed energies are marked with "#"; otherwise, they are calculated in this work with Cowan's codes (Cowan 1981).

[e] Comments to transition identifications: A – we added additional components to the classification; C – observed in photon emission by Cantù et al. (1981); N – new classification; and R – the original identification was revised in this work.

## 3. OPTIMIZATION OF ENERGY LEVELS

Once the list of observed wavelengths of identified transitions and their estimated uncertainties was constructed, it was inserted into the least squares level optimization process that uses a computer code LOPT (Kramida 2011). In this process, consistency of line classifications, observed wavelengths, and their uncertainties can further be checked by comparison with the Ritz wavelength. A detailed description of methodology of such analysis can be found in previous NIST publications (Kramida 2013a, 2013b, 2013c; Haris et al. 2014). In case of C I, the level optimization was started with laboratory FT measurements and a few other high-precision measurements of transitions terminating on the ground levels. Deviating lines, most of which were blended, asymmetric, or multiply classified, were further checked, and either their wavelengths were remeasured or their uncertainties were revised. For many lines, uncertainties were increased after methodical examination of their profile. Contributions of perturbing nearby lines could often be detected by analyzing the line intensity, observed width, or shape. For asymmetric lines, the adopted uncertainty is the difference in their centroid value obtained by two methods, by least squares fitting of a Voigt profile and by determining the center of gravity. Once all excited levels determined from long-wavelength lines have stabilized, the Ritz wavelengths based on them were used to internally calibrate less accurate measurements and assess their uncertainties. A systematic correction was needed for only one previously reported observation, that of Kaufman & Ward (1966) (see Section 2.4). When multiple wavelength measurements were available for a given transition, the most accurate one was kept in level optimization; only those best-measured wavelengths are inserted in the list of classified lines (Table 2). Nevertheless, intensities from all available observations were taken into account to derive an average intensity of each line (see Section 5 below).

Presentation of uncertainties of atomic energy levels is a non-trivial problem. On the one hand, excitation energies are widely used quantities, and their uncertainties can be well-defined. On the other hand, uncertainties of excitation energies are often much larger than uncertainties of separations between energy levels, which are defined by measurements of transition wavelengths. In many atomic spectra, and C I is one of them, wavelengths of transitions between excited levels can be measured much more accurately than those of transitions to the ground level. In such situation, it is often possible to select one of the excited levels as a "base" level for the determination of relative uncertainties, so that a combination in quadrature of those relative uncertainties provides a good estimate of uncertainty for wavenumbers of most transitions. In C I, we chose the $2s^22p3p\ ^3P_2$ level at 71 385.40 992 $cm^{-1}$ as such a base level (BL). This level has one of the smallest uncertainties of separations from other levels, $D_1$ = 0.00006 $cm^{-1}$ (for definition of $D_1$, see Section 2.1), and one of the largest numbers of observed spectral lines originating from or terminating on this level (43). Since high-precision measurements were made in several fragmented regions of the spectrum, it is practically impossible to give a single set of uncertainties that would perfectly describe all energy levels. This is the reason for the presence of an additional column of the $D_1$ uncertainties in Table 2. Levels having non-empty $D_1$ values have more significant digits than would be justified by their uncertainty relative to the BL. With our choice of BL, a combination in quadrature of the uncertainties (of separations from BL) provides a good estimate of uncertainty for the vast majority of Ritz wavenumbers of all possible transitions. This is confirmed by comparing this uncertainty estimate with the accurate wavenumber uncertainty computed by the LOPT code with account for covariance of the optimized levels (see Kramida 2011). Among the 2102 transitions included in Table 1, only about 50 have the Ritz wavenumber uncertainty smaller by more than a factor of two than the one estimated from a combination in quadrature of the corresponding level uncertainties. The marked exceptions are the forbidden transitions within the ground term, for which the above difference is a factor of a few thousands. In general, if one wants to estimate an uncertainty of a Ritz wavenumber of any transition not included in Table 1, one should look at the $D_1$ values of the energy levels involved. If both are blank, the combination in quadrature of the values in column "Unc." of Table 2 is a good estimate of uncertainty, accurate to within a factor of two. If any of them are non-blank, such an estimate is likely too large by a factor of up to six.

## 4. ISOTOPIC SHIFTS AND HYPERFINE STRUCTURES

Natural carbon samples contain two stable isotopes, $^{12}$C (98.94 %, nuclear moment $I$ = 0) and $^{13}$C (1.06 %, $I$ = 1/2), and traces of unstable (cosmogenic) $^{14}$C (< $10^{-4}$ %, $I$ = 0, half-life $\tau_{½}$ = 5700(30) years) (Meija et al. 2016). In addition, many artificial isotopes from $^{8}$C to $^{22}$C were identified, among them $^{10}$C ($I$ = 0, $\tau_{½}$ = 19.308(4) s) and $^{11}$C ($I$ = 3/2, $\tau_{½}$ = 20.334(24) min) are the longest- lived (Sonzogni 2016). In neutral carbon, isotopic shifts (IS) have been measured for only a few transitions, but many more have been accurately calculated. These data (described below) allowed us to derive accurate values for transition wavelengths and energy levels of carbon isotopes, which are collected in Tables 6 and 7. As it is common in the literature on IS, we use a unit of millikayser (1 mK = $10^{-3}$ cm$^{-1}$) in this Section and relevant tables.

Since the determination of the IS of the ground-term levels is largely based on the high-precision measurements of hyperfine structures in $^{13}$C (Yamamoto & Saito 1991; Klein et al. 1998, Wolber et al. 1970), we start with the discussion of these measurements. The relevant data are collected in Table 8. It should be noted that both Yamamoto & Saito and Klein et al. give uncertainties on the level of three standard deviations, while Wolber et al. reported the uncertainty of the observed HF splitting on the level of "slightly more than one standard deviation" for the $2s^2 2p^2\ {}^3P_1$ level and two standard deviations for ${}^3P_2$. In Table 8, we reduced all the reported uncertainties to the level of one standard deviation, assuming normal statistical distribution for all these measurements. Thus, the results of Wolber et al. are adopted here as 4.200(25) MHz and 372.593(13) MHz for the HF splitting of the ${}^3P_1$ and ${}^3P_2$ levels, respectively. A combination of those results with the absolute frequencies of the HF components of the ${}^3P_0$–${}^3P_1$ and ${}^3P_1$–${}^3P_2$ transitions (two components were measured for each) reported by Yamamoto & Saito and Klein et al. yields six input data values for the least-squares determination of the four HF levels of ${}^{13}$C, ${}^3P_1$ ($F$ = 1/2, 3/2) and ${}^3P_2$ ($F$ = 3/2, 5/2). We optimized the levels using the LOPT code (Kramida 2011). As expected, uncertainties of the optimized levels and HF intervals are slightly smaller than those of the measurements, and the optimized values agree with measurements within the uncertainties. The frequencies of the fine-structure transitions in ${}^{13}$C given in Table 8 were determined as weighted means of the Ritz wavenumbers of the corresponding hyperfine transitions with weights equal to their theoretical relative intensities.

To complete the discussion of the above measurements, we note that the magnetic dipole constants $A_{\mathrm{hfs}}$ of the hyperfine splitting in ${}^{13}$C and Landé $g_J$ factors of ${}^3P_{1,2}$ of ${}^{12}$C were accurately measured by Wolber et al. (1970, 1969). The $A_{\mathrm{hfs}}$ values reported by Wolber et al. (1970) for the ${}^3P_1$ and ${}^3P_2$ levels of ${}^{13}$C are +2.838(17) MHz and +149.055(10) MHz (the uncertainty in the latter value represents two standard deviations), and the $g_J$ factors for the same levels of ${}^{12}$C are 1.501052(13) and 1.501039(15), respectively (Wolber et al. 1969). This helped those authors to resolve an ambiguity in the HF splitting constant of the $^3P_1$ state of the exotic $^{11}$C isotope measured by Haberstroh et al. (1964), yielding $A_{\rm hfs}$ ($^{11}$C, $^3P_1$) = −1.308(24) MHz.

In addition to the HF ground-term transitions in $^{13}$C, Yamamoto & Saito (1991) and Klein et al. (1998) measured the frequencies of the corresponding fine-structure transitions in $^{12}$C. Their results are 492 160.651(18) MHz and 809 341.970(17) MHz for the $^{12}$C $^3P_0$–$^3P_1$ and $^3P_1$–$^3P_2$ transitions, respectively. Combined with our optimized values for the centers of gravity of those transitions in $^{13}$C (see Table 8), this yields the IS($^{13}$C–$^{12}$C) of these transitions to be +0.0747(8) mK and +0.1378(16) mK, respectively. Positive shifts mean that the wavelengths of heavy isotopes are shifted toward shorter wavelengths, similar to those of a one-electron atom.

Such positive IS was also observed for transitions from the $2s2p^3$ levels. In particular, a shift of +670(5) mK was observed for the transition from $2s2p^3$ $^5S°_2$ to the ground-term level $^3P_2$ (Bernheim & Kittrell 1980). Similarly, a positive IS was observed for all three $2s^22p^2$ $^3P_{0,1,2}$ – $2s2p^3$ $^3S°_1$ transitions near 945 Å (Labazan et al. 2005). The latter authors also made an absolute measurement of the frequencies of all the $^{12}$C and $^{13}$C components (see Table 6). In their separate IS measurements, systematic uncertainties were largely canceled out. They succeeded to measure the IS of two fine-structure components (from the $^3P_0$ and $^3P_2$ levels of $2s^22p^2$), while the $^3P_1$ component could not be measured due to experimental constraints. Nevertheless, it could be determined, albeit with lower accuracy, from the absolute measurements. Combining the data for the three transitions, they derived a value of +510.7(13) mK for the weighted mean of the three transitions. In our level optimization procedure described below, we made use of only the two direct measurements of the IS from that work, which give a weighted mean of 510.9(15) mK. It should be noted that in Table II of Labazan et al. (2005) an incorrect value for the speed of light (exactly $3\times10^{10}$ cm/s) was used to convert the correctly given IS values in units of MHz to those in $cm^{-1}$ (Ubachs 2016). In our work, we used the original values in MHz from that table.

The IS of the $2s^22p^2\ ^1S_0 - 2s^22p3s\ ^1P°_1$ transition near 2479 Å, which turned out to be negative, was measured for all three natural isotopes. It is −156(2) mK and −294(2) mK, for $^{13}C$–$^{12}C$ and $^{14}C$–$^{12}C$, respectively (Holmes 1951), and the first one agrees with a previous measurement by Burnett (1950). Using the above shifts and Ritz wavelengths of natural C (see Table 1), we determined the wavelengths of isotopes 12 through 14 to be 2478.561 30(19) Å, 2478.570 89(22) Å and 2478.579 37(22) Å, respectively.

The experimental IS data described above, when combined with Ritz wavelengths and energy levels for natural carbon, are sufficient to accurately determine the absolute positions of the $2s^22p^2\ ^3P_{1,2}$, $2s2p^3\ ^5S°_2$, and $2s2p^3\ ^3S°_1$ levels in both $^{12}C$ and $^{13}C$. However, to precisely locate the $2s^22p^2\ ^1S_0$ and $2p3s\ ^1P°_1$ levels, additional data on IS of these levels are needed. We have used the calculated IS($^{13}C$–$^{12}C$) for the $2p3s\ ^1P°$ level, 116(4) mK (Berengut et al. 2006) in combination with the IS of the $2s^22p^2\ ^1S_0 - 2s^22p3s\ ^1P°_1$ transition measured by Holmes (1951), which yields +40(5) mK for the IS($^{13}C$–$^{12}C$) of the $2s^22p^2\ ^1S_0$ level. This provided enough data to make a least-squares level optimization with the same LOPT code (Kramida 2011) and determine those levels and Ritz wavelengths of transitions between them for both $^{12}C$ and $^{13}C$.

The paper of Berengut et al. (2006) provides calculated IS($^{13}C$–$^{12}C$) values for 28 energy levels with an estimated uncertainty of 4 mK. We note that, although there are several different statements in that paper regarding the uncertainties, those authors prefer the value given above (Kozlov 2016). By analyzing the comparisons with other calculations and measurements given in various tables of Berengut et al., we verified that this uncertainty estimate is statistically consistent with other published data. The IS of one level of the ground configuration, $2s^22p^2\ ^1D_2$, which was not included in Berengut et al. (2006), was calculated by Kozlov et al. (2009) with somewhat lower accuracy estimated as 9 mK (Kozlov 2016).

From experimental and theoretical data discussed above, combined with the Ritz wavelengths and energy levels of natural C given in Tables 1 and 2, we derived the transition wavelengths and energy levels of the $^{12}$C and $^{13}$C isotopes given in Tables 6 and 7. Wavenumbers of seventeen lines of several low-lying multiplets at (1329, 1561, 1657 and 1931) Å in $^{13}$C were measured by Haridass & Huber (1994) with an estimated uncertainty of 0.1 cm$^{-1}$. These measurements agree with the data compiled in Table 6 and 7 within the quoted uncertainty. The IS($^{14}$C–$^{12}$C) can easily be derived from these data by scaling the values for IS($^{13}$C–$^{12}$C) with a factor of 1.853 98, assuming that the IS for each pair of isotopes is proportional to the difference of inverse atomic masses of the two isotopes. Thus, calculated IS($^{14}$C–$^{12}$C) of the $2s^2 2p^2\ ^1S_0 - 2s^2 2p3s\ ^1P°_1$ transition is −289(4) mK, in fair agreement with the measurement of Holmes (1951), −294(2) mK.

## Table 6

Frequencies, Wavelengths, and Isotopic Shifts of Selected Transitions in $^{12}$C and $^{13}$C

| Transition [a] | $^{12}$C I [b] | $^{13}$C I [b] | IS($^{13}$C–$^{12}$C)[c] | | Ref. [d] |
|---|---|---|---|---|---|
| | | | TW | Others | |
| $^3P_0 - {}^3P_1$ | *492160.651(18)* | *492162.889(18)* | 0.0747(8) | 0.0750(10) | Y91 |
| $^3P_1 - {}^3P_2$ | *809341.970(17)* | *809346.103(44)* | 0.1378(16) | 0.138(3) | K98 |
| $^3P_0$–$^1D_2$ | [9850.250(3)] | [9850.245(9)] | | (5(9)) | K09 |
| $^3P_1$–$^1D_2$ | [9824.118(3)] | [9824.113(9)] | | (5(9)) | K09 |
| $^3P_2$–$^1D_2$ | [9808.295(3)] | [9808.290(9)] | | (5(9)) | K09 |
| $^1D_2$–$^1S_0$ | [8727.131(3)] | [8727.108(8)] | | (30(10)) | K09,B06 |
| $^3P_2$–$^1S_0$ | [4627.3445(6)] | [4627.3369(11)] | | (35(4)) | B06 |
| $^3P_1$–$^1S_0$ | [4621.5694(6)] | [4621.5618(11)] | | (35(4)) | B06 |
| $^1D_2$–$2s2p^3\ {}^5S^\circ_2$ | [4246.4495(28)] | [4246.3296(32)] | | (669(10)) | K09,B80 |
| $^3P_2$–$2s2p^3\ {}^5S^\circ_2$ | 2967.2236(14) | 2967.1646(15) | | 670(5) | B80 |
| $^3P_1$–$2s2p^3\ {}^5S^\circ_2$ | 2964.8478(14) | 2964.7889(15) | | | |
| $^1S_0$–$2s^22p3s\ {}^3P^\circ_1$ | [2582.89721(22)] | [2582.9057(5)] | | (−127(6)) | B06 |
| $^1S_0$–$2s^22p3s\ {}^1P^\circ_1$ | [2478.56131(19)] | [2478.57089(30)] | | −156(2) | H51 |
| $^1D_2$–$2s^22p3s\ {}^3P^\circ_1$ | [1993.62033(13)] | [1993.6242(4)] | | (−97(10)) | K09,B06 |
| $^1D_2$–$2s^22p3s\ {}^3P^\circ_2$ | [1992.01151(13)] | [1992.0154(4)] | | (−97(10)) | K09,B06 |
| $^1D_2$–$2s^22p3s\ {}^1P^\circ_1$ | [1930.90540(12)] | [1930.9099(4)] | | (−121(10)) | K09,B06 |
| $^1S_0$–$2s^22p4s\ {}^3P^\circ_1$ | [1770.89117(10)] | [1770.89314(21)] | | (−63(6)) | B06 |
| $^1S_0$–$2s^22p3d\ {}^3D^\circ_1$ | [1765.36582(10)] | [1765.36790(21)] | | (−67(6)) | B06 |
| $^1S_0$–$2s^22p4s\ {}^1P^\circ_1$ | [1763.90896(10)] | [1763.91111(21)] | | (−69(6)) | B06 |
| $^1S_0$–$2s^22p3d\ {}^1P^\circ_1$ | [1751.82688(10)] | [1751.82891(21)] | | (−66(6)) | B06 |
| $^1S_0$–$2s^22p3d\ {}^3P^\circ_1$ | [1733.98039(10)] | [1733.97961(21)] | | (26(6)) | B06 |
| $^3P_2$–$2s^22p3s\ {}^3P^\circ_1$ | [1658.12056(4)] | [1658.12308(12)] | | (−92(4)) | B06 |
| $^3P_1$–$2s^22p3s\ {}^3P^\circ_0$ | [1657.90652(4)] | [1657.90904(12)] | | (−92(4)) | B06 |
| $^3P_1$–$2s^22p3s\ {}^3P^\circ_1$ | [1657.37865(4)] | [1657.38117(12)] | | (−92(4)) | B06 |
| $^3P_2$–$2s^22p3s\ {}^3P^\circ_2$ | [1657.00751(4)] | [1657.01004(12)] | | (−92(4)) | B06 |
| $^3P_0$–$2s^22p3s\ {}^3P^\circ_1$ | [1656.92782(4)] | [1656.93034(12)] | | (−92(4)) | B06 |
| $^3P_1$–$2s^22p3s\ {}^3P^\circ_2$ | [1656.26660(4)] | [1656.26912(12)] | | (−92(4)) | B06 |
| $^3P_2$–$2s^22p3s\ {}^1P^\circ_1$ | [1614.50680(3)] | [1614.50982(12)] | | (−116(4)) | B06 |
| $^3P_1$–$2s^22p3s\ {}^1P^\circ_1$ | [1613.80340(3)] | [1613.80642(12)] | | (−116(4)) | B06 |
| $^3P_0$–$2s^22p3s\ {}^1P^\circ_1$ | [1613.37596(3)] | [1613.37898(12)] | | (−116(4)) | B06 |
| $^3P_2$–$2s2p^3\ {}^3D^\circ_3$ | [1561.43753(3)] | [1561.42037(11)] | | (704(4)) | B06 |
| $^3P_2$–$2s2p^3\ {}^3D^\circ_1$ | [1561.36611(3)] | [1561.34896(11)] | | (704(4)) | B06 |
| $^3P_2$–$2s2p^3\ {}^3D^\circ_2$ | [1561.33943(3)] | [1561.32228(11)] | | (704(4)) | B06 |
| $^3P_1$–$2s2p^3\ {}^3D^\circ_1$ | [1560.70824(3)] | [1560.69110(11)] | | (704(4)) | B06 |
| $^3P_1$–$2s2p^3\ {}^3D^\circ_2$ | [1560.68158(3)] | [1560.66444(11)] | | (704(4)) | B06 |
| $^3P_0$–$2s2p^3\ {}^3D^\circ_1$ | [1560.30846(3)] | [1560.29133(11)] | | (704(4)) | B06 |
| $^1D_2$–$2s^22p3d\ {}^1D^\circ_2$ | [1481.76299(7)] | [1481.76335(22)] | | (−17(10)) | K09,B06 |
| $^1D_2$–$2s^22p4s\ {}^3P^\circ_1$ | [1472.23115(7)] | [1472.23185(22)] | | (−32(10)) | K09,B06 |
| $^1D_2$–$2s^22p4s\ {}^3P^\circ_2$ | [1471.55229(7)] | [1471.55303(22)] | | (−34(10)) | K09,B06 |
| $^1D_2$–$2s^22p3d\ {}^3F^\circ_2$ | [1470.44909(7)] | [1470.44964(22)] | | (−25(10)) | K09,B06 |
| $^1D_2$–$2s^22p3d\ {}^3F^\circ_3$ | [1470.09352(7)] | [1470.09408(22)] | | (−26(10)) | K09,B06 |

| Transition [a] | $^{12}$C I [b] | $^{13}$C I [b] | IS($^{13}$C–$^{12}$C)[c] | | Ref. [d] |
|---|---|---|---|---|---|
| | | | TW | Others | |
| $^1D_2$–$2s^22p3d\ ^3D°_1$ | [1468.41033(7)] | [1468.41111(22)] | | (−36(10)) | K09,B06 |
| $^1D_2$–$2s^22p3d\ ^3D°_2$ | [1468.10564(7)] | [1468.10641(22)] | | (−36(10)) | K09,B06 |
| $^1D_2$–$2s^22p3d\ ^3D°_3$ | [1467.87678(7)] | [1467.87755(22)] | | (−36(10)) | K09,B06 |
| $^1D_2$–$2s^22p4s\ ^1P°_1$ | [1467.40222(7)] | [1467.40306(22)] | | (−39(10)) | K09,B06 |
| $^1D_2$–$2s^22p3d\ ^1F°_3$ | [1463.33633(7)] | [1463.33718(21)] | | (−39(10)) | K09,B06 |
| $^1D_2$–$2s^22p3d\ ^1P°_1$ | [1459.03102(7)] | [1459.03178(21)] | | (−36(10)) | K09,B06 |
| $^1D_2$–$2s^22p4d\ ^1D°_2$ | [1364.16474(6)] | [1364.16441(19)] | | (18(10)) | K09,B06 |
| $^3P_2$–$2s2p^3\ ^3P°_1$ | [1329.600133(23)] | [1329.59017(8)] | | (564(4)) | B06 |
| $^3P_2$–$2s2p^3\ ^3P°_2$ | [1329.577293(23)] | [1329.56733(8)] | | (564(4)) | B06 |
| $^3P_1$–$2s2p^3\ ^3P°_1$ | [1329.123046(23)] | [1329.11308(8)] | | (564(4)) | B06 |
| $^3P_1$–$2s2p^3\ ^3P°_2$ | [1329.100222(23)] | [1329.09026(8)] | | (564(4)) | B06 |
| $^3P_1$–$2s2p^3\ ^3P°_0$ | [1329.084750(23)] | [1329.07479(8)] | | (564(4)) | B06 |
| $^3P_0$–$2s2p^3\ ^3P°_1$ | [1328.833097(23)] | [1328.82314(8)] | | (564(4)) | B06 |
| $^3P_2$–$2s^22p3d\ ^1D°_2$ | [1288.055276(22)] | [1288.05547(7)] | | (−12(4)) | B06 |
| $^3P_1$–$2s^22p3d\ ^1D°_2$ | [1287.607533(22)] | [1287.60772(7)] | | (−12(4)) | B06 |
| $^3P_2$–$2s^22p4s\ ^3P°_1$ | [1280.846628(21)] | [1280.84708(7)] | | (−28(4)) | B06 |
| $^3P_1$–$2s^22p4s\ ^3P°_0$ | [1280.596917(21)] | [1280.59735(7)] | | (−26(4)) | B06 |
| $^3P_1$–$2s^22p4s\ ^3P°_1$ | [1280.403881(21)] | [1280.40433(7)] | | (−27(4)) | B06 |
| $^3P_2$–$2s^22p4s\ ^3P°_2$ | [1280.332764(21)] | [1280.33325(7)] | | (−30(4)) | B06 |
| $^3P_0$–$2s^22p4s\ ^3P°_1$ | [1280.134796(21)] | [1280.13524(7)] | | (−27(4)) | B06 |
| $^3P_1$–$2s^22p4s\ ^3P°_2$ | [1279.890372(21)] | [1279.89085(7)] | | (−29(4)) | B06 |
| $^3P_2$–$2s^22p3d\ ^3F°_2$ | [1279.497567(21)] | [1279.49790(7)] | | (−21(4)) | B06 |
| $^3P_2$–$2s^22p3d\ ^3F°_3$ | [1279.228340(21)] | [1279.22869(7)] | | (−21(4)) | B06 |
| $^3P_1$–$2s^22p3d\ ^3F°_2$ | [1279.055753(21)] | [1279.05609(7)] | | (−20(4)) | B06 |
| $^3P_2$–$2s^22p3d\ ^3D°_1$ | [1277.953648(21)] | [1277.95416(7)] | | (−32(4)) | B06 |
| $^3P_2$–$2s^22p3d\ ^3D°_2$ | [1277.722865(21)] | [1277.72337(7)] | | (−31(4)) | B06 |
| $^3P_2$–$2s^22p3d\ ^3D°_3$ | [1277.549512(21)] | [1277.55002(7)] | | (−31(4)) | B06 |
| $^3P_1$–$2s^22p3d\ ^3D°_1$ | [1277.512899(21)] | [1277.51341(7)] | | (−31(4)) | B06 |
| $^3P_1$–$2s^22p3d\ ^3D°_2$ | [1277.282274(21)] | [1277.28278(7)] | | (−31(4)) | B06 |
| $^3P_0$–$2s^22p3d\ ^3D°_1$ | [1277.245028(21)] | [1277.24554(7)] | | (−31(4)) | B06 |
| $^3P_2$–$2s^22p4s\ ^1P°_1$ | [1277.190025(21)] | [1277.19058(7)] | | (−34(4)) | B06 |
| $^3P_1$–$2s^22p4s\ ^1P°_1$ | [1276.749802(21)] | [1276.75035(7)] | | (−34(4)) | B06 |
| $^3P_0$–$2s^22p4s\ ^1P°_1$ | [1276.482251(21)] | [1276.48280(7)] | | (−34(4)) | B06 |
| $^3P_2$–$2s^22p3d\ ^1F°_3$ | [1274.108793(21)] | [1274.10936(7)] | | (−35(4)) | B06 |
| $^3P_2$–$2s^22p3d\ ^1P°_1$ | [1270.843705(21)] | [1270.84420(7)] | | (−31(4)) | B06 |
| $^3P_1$–$2s^22p3d\ ^1P°_1$ | [1270.407846(21)] | [1270.40834(7)] | | (−31(4)) | B06 |
| $^3P_0$–$2s^22p3d\ ^1P°_1$ | [1270.142946(21)] | [1270.14344(7)] | | (−31(4)) | B06 |
| $^3P_2$–$2s^22p3d\ ^3P°_2$ | [1261.551809(21)] | [1261.55081(7)] | | (63(4)) | B06 |
| $^3P_2$–$2s^22p3d\ ^3P°_1$ | [1261.425438(21)] | [1261.42446(7)] | | (61(4)) | B06 |
| $^3P_1$–$2s^22p3d\ ^3P°_2$ | [1261.122299(21)] | [1261.12130(7)] | | (63(4)) | B06 |
| $^3P_1$–$2s^22p3d\ ^3P°_1$ | [1260.996014(21)] | [1260.99504(7)] | | (61(4)) | B06 |
| $^3P_1$–$2s^22p3d\ ^3P°_0$ | [1260.926387(22)] | [1260.92542(7)] | | (61(4)) | B06 |
| $^3P_0$–$2s^22p3d\ ^3P°_1$ | [1260.735024(21)] | [1260.73405(7)] | | (61(4)) | B06 |
| $^3P_2$–$2s^22p4d\ ^1D°_2$ | [1198.262427(19)] | [1198.26211(6)] | | (22(4)) | B06 |

| Transition [a] | $^{12}$C I [b] | $^{13}$C I [b] | IS($^{13}$C–$^{12}$C)[c] | | Ref. [d] |
|---|---|---|---|---|---|
| | | | TW | Others | |
| $^3P_1$–$2s^22p4d\ ^1D°_2$ | [1197.874924(19)] | [1197.87460(6)] | | (22(4)) | B06 |
| $^3P_0$–$2s2p^3\ ^3S°_1$ | 945.18750(3) | 945.18294(3) | | | |
| $^3P_1$–$2s2p^3\ ^3S°_1$ | 945.33419(3) | 945.32962(3) | 511.0(15) | 510.7(13) | L05 |
| $^3P_2$–$2s2p^3\ ^3S°_1$ | 945.57551(3) | 945.57094(3) | | | |

[a] The ground-term levels $2s^22p^2\ ^ML_J$ are abbreviated as $^ML_J$.

[b] Observed wavelength of spectral lines of isotopes $^{12}$C and $^{13}$C in Å (in standard air above 2000 Å, in vacuum below that), except the first two transitions, for which the energy in MHz is given in italics. The uncertainty (at the 1-sigma level) in the last digit is given in parentheses. The values without square brackets are Ritz wavelength resulting from our least-squares optimization (see text). The values in square brackets are derived partially from theoretical data.

[c] The IS referred to this work (TW) is derived from our Ritz wavelengths. Measured or theoretical values appear in the next column. Theoretical values are enclosed in parentheses. All IS values are given in millikaysers (1 mK = $10^{-3}$ cm$^{-1}$).

[d] References for measured values in columns 2, 3, and 5: B06 – Berengut et al. (2006); B80 – Bernheim & Kittrell (1980); H51 – Holmes (1951); K09 – Kozlov et al. (2009); K98 – Klein et al. (1998); L05 – Labazan et al. (2005); Y91 – Yamamoto & Saito (1991).

## Table 7

### Observed and Predicted Energy Levels of Isotopes $^{12}$C and $^{13}$C

| Level designation | $E$($^{12}$C) (cm$^{-1}$) | $E$($^{13}$C) (cm$^{-1}$) | IS($^{12}$C–$^{13}$C) [a] (mK) |
|---|---|---|---|
| $2s^22p^2\ ^3P_0$ | 0.0000000 | 0.0000000 | – |
| $2s^22p^2\ ^3\mathrm{P}_1$ | 16.4167122(6) | 16.4167869(6) | +0.0747(8) |
| $2s^22p^2\ ^3P_2$ | 43.4134544(8) | 43.4136669(16) | +0.2125(18) |
| $2s^22p^2\ ^1D_2$ | [10192.657(3)][b] | [10192.662(9)][b] | (+5(9))[b] |
| $2s^22p^2\ ^1S_0$ | [21648.030(3)][c] | [21648.070(4)][c] | [(+40(4))][c] |
| $2s2p^3\ ^5S°_2$ | 33735.114(16) | 33735.784(17) | +670(5) |
| $2s^22p3s\ ^3P°_0$ | [60333.4484(14)] | [60333.357(3)] | (−92(3)) |
| $2s^22p3s\ ^3P°_1$ | [60352.6592(13)] | [60352.567(3)] | (−92(3)) |
| $2s^22p3s\ ^3P°_2$ | [60393.1702(13)] | [60393.078(3)] | (−92(3)) |
| $2s^22p3s\ ^1P°_1$ | [61981.8333(13)][c] | [61981.7173(24)][c] | (−116(4)) |
| $2s2p^3\ ^3D°_3$ | [64086.9620(14)] | [64087.666(5)] | (+704(5)) |
| $2s2p^3\ ^3D°_1$ | [64089.8914(15)] | [64090.595(5)] | (+704(5)) |
| $2s2p^3\ ^3D°_2$ | [64090.9859(14)] | [64091.690(5)] | (+704(5)) |
| $2s2p^3\ ^3P°_1$ | [75253.9956(13)] | [75254.560(3)] | (+564(3)) |
| $2s2p^3\ ^3P°_2$ | [75255.2876(13)] | [75255.851(3)] | (+564(3)) |
| $2s2p^3\ ^3P°_0$ | [75256.1635(13)] | [75256.727(3)] | (+564(3)) |
| $2s^22p3d\ ^1D°_2$ | [77679.8331(13)] | [77679.822(3)] | (−11(3)) |
| $2s^22p4s\ ^3P°_0$ | [78105.0008(13)] | [78104.974(3)] | (−26(3)) |
| $2s^22p4s\ ^3P°_1$ | [78116.7735(13)] | [78116.746(3)] | (−27(3)) |
| $2s^22p4s\ ^3P°_2$ | [78148.1084(13)] | [78148.079(3)] | (−29(3)) |
| $2s^22p3d\ ^3F°_2$ | [78199.0915(13)] | [78199.071(3)] | (−20(3)) |
| $2s^22p3d\ ^3F°_3$ | [78215.5402(13)] | [78215.519(3)] | (−21(3)) |
| $2s^22p3d\ ^3D°_1$ | [78293.5129(13)] | [78293.481(3)] | (−31(3)) |
| $2s^22p3d\ ^3D°_2$ | [78307.6465(13)] | [78307.616(3)] | (−31(3)) |
| $2s^22p3d\ ^3D°_3$ | [78318.2662(13)] | [78318.235(3)] | (−31(3)) |
| $2s^22p4s\ ^1P°_1$ | [78340.2981(13)] | [78340.264(3)] | (−34(3)) |
| $2s^22p3d\ ^1F°_3$ | [78529.6468(13)] | [78529.612(3)] | (−34(3)) |
| $2s^22p3d\ ^1P°_1$ | [78731.2958(13)] | [78731.265(3)] | (−31(3)) |
| $2s^22p3d\ ^3P°_2$ | [79310.8674(13)] | [79310.931(3)] | (+63.1(22)) |
| $2s^22p3d\ ^3P°_1$ | [79318.8085(13)] | [79318.870(3)] | (+61.4(22)) |
| $2s^22p3d\ ^3P°_0$ | [79323.1875(14)] | [79323.248(3)] | (+60.8(22)) |
| $2s^22p4d\ ^1D°_2$ | [83497.5865(13)] | [83497.609(3)] | (+23(3)) |
| $2s2p^3\ ^3S°_1$ | 105799.1137(30) | 105799.6247(30) | +511.0(15) |

[a] Isotopic level shift IS = $E_{obs}^{13} - E_{obs}^{12}$ in millikaysers (1 mK = $10^{-3}$ cm$^{-1}$) derived from our level optimization (see text).

[b] Positions of these levels was determined from observed energy levels of natural C (see Table 2) and isotope shifts calculated by Kozlov et al. (2009). All other levels in square brackets, unless otherwise stated, were determined similarly, using isotope shifts calculated by Berengut et al. (2006).

[c] The position of the $^{12}$C $2s^22p3s\ ^1P°_1$ level was fixed in the level optimization at the value calculated from the optimized value for natural C (see Table 2) and the IS of this level given in the last column, quoted from Berengut et al. (2006).

## Table 8

Frequencies of Selected Hyperfine Transitions Within the Ground Term of $^{13}$C I (MHz)

| Transition | $I_{calc}$[a] | Transition Frequency | | Center of gravity[b] | Ref.[c] |
|---|---|---|---|---|---|
| | | Observed | Ritz | | |
| $^3P_1$ | | | | | |
| $F$ = 1/2–3/2 | | 4.200(25) | 4.195(23) | | W70 |
| $^3P_2$ | | | | | |
| $F$ = 3/2–5/2 | | 372.593(13) | 372.591(13) | | W70 |
| $^3P_0 - {}^3P_1$ | | | | 492162.900(35) | Y91 |
| | | | | 492162.889(18) | TW |
| $F$ = 1/2–1/2 | 0.333 | 492160.147(56) | 492160.091(29) | | Y91 |
| $F$ = 1/2–3/2 | 0.667 | 492164.276(24) | 492164.286(22) | | Y91 |
| $^3P_1 - {}^3P_2$ | | | | 809346.1(1) | K98 |
| | | | | 809346.103(44) | TW |
| $F$ = 1/2–3/2 | 0.333 | 809125.50(13) | 809125.346(66) | | K98 |
| $F$ = 3/2–3/2 | 0.067 | 809121.30(13)[d] | 809121.152(63) | | K98 |
| $F$ = 3/2–5/2 | 0.600 | 809493.70(7) | 809493.743(63) | | K98 |

[a] Calculated relative intensities of HF components.

[b] Center of gravity of transitions between fine-structure levels determined from the observed and Ritz values (first and second entries, respectively).

[c] References for the values in columns 3 and 5: K98 – Klein et al. (1998); Y91 – Yamamoto & Saito (1991); W70 – Wolber et al. (1970); TW – determined by this work either from observed or Ritz.

[d] Calculated from the measured $^3P_{1,F=1/2}$–$^3P_{2,F=3/2}$ frequency and the $^3P_{1,F=1/2}$–$^3P_{1,F=3/2}$ separation reported by Wolber et al. (1970).

## 5. REDUCTION OF OBSERVED INTENSITIES

Different observers used grossly different scales for relative line intensities. Reducing multiple observations for all the observed lines to a common uniform scale is a non-trivial problem, which was well addressed in analyses of many laboratory spectra (Kramida 2013a, 2013b, 2013c; Haris et al. 2014). In this method, the effective excitation temperature ($T_{\mathrm{eff}}^{\mathrm{exc}}$) is derived for a given atom or ion within a simplified model assuming a local thermodynamic equilibrium (LTE) describing the level populations by Maxwell-Boltzmann equations for an optically thin source. The optical response function of the instrument (of both optics and registration equipment) is derived from the ratios of calculated and observed intensities. This instrumental function is subtracted from the observed intensities before deriving the more accurate $T_{\mathrm{eff}}^{\mathrm{exc}}$ from the Boltzmann plot. The procedure is repeated in several iterations until the convergence is achieved. Once the effective temperatures of all observations are obtained, the observed intensities can easily be converted to a uniform scale corresponding to a common excitation temperature. When observations are available from many sources, we averaged those converted intensities from each of them. If the scaled intensities from different sources varied by more than an order of magnitude, those observations were further checked and either corrected or dropped from averaging.

Apart from observed intensities and wavelengths, important required parameters for the above process are the weighted transition rates (*gA*). For those, we used either critically evaluated values from Wiese et al. (1996) and (Wiese & Fuhr 2007) or our own evaluated data computed with Cowan's codes (Cowan 1981) (see Section 6). In this work, we chose the effective temperature derived from the 85R13 FT spectrum as the basis for the global intensity scale, to which we reduced the modeled intensities of all other measurements. For that spectrum, we derived $T_{\mathrm{eff}}^{\mathrm{exc}} = 0.41$ eV from 287 lines selected out of total 319 lines observed in it.

## 6. THEORETICAL CALCULATIONS

We used our calculations with the Windows-PC version of Cowan's codes (Cowan 1981), made in frames of the superposition-of-configurations Hartree-Fock formalism, to provide theoretical support for the observed energy levels and wavelengths and to calculate the transition rates. Extensive calculations were made for configurations of both parities. The even-parity set comprised configurations $2s^22pn\ell$ ($n \leq 13$, 10, and 9 for $\ell = p, f$, and $h$, respectively) and $2s2p^2n\ell$ ($n \leq 5$ and 4 for $\ell = s$ and $d$, respectively), while the odd-parity set included $2s^22pn\ell$ ($n \leq 21$ and 10, for $\ell = s$+$d$ and $g$, respectively) and $2s2p^2n\ell$ ($n \leq 5$ for $\ell = p$+$f$). The initial (scaled *ab initio*) values of the Slater parameters were adjusted to obtain a least-squares fit (LSF) of calculated energy levels to their observed values. These fitted parameters were further used to calculate improved transition parameters. In the LSF, the even and odd sets of levels were fitted with a standard deviation of 40 $cm^{-1}$ and 23 $cm^{-1}$, respectively. The parameters obtained in the LSF are summarized in Table 9 (available only as a supplementary online material).

Most of the transition parameters, such as transition rates $A_{ki}$, oscillator strengths $f_{ik}$, and line strengths $S_{ik}$, were taken from various theoretical calculations. Only a small number of experimental data for these parameters of C I exist in the literature. The compilation and assessment of such a large amount of data could be a subject of a separate project. However, for neutral carbon, such an assessment was recently made by Wiese et al. (1996) and Wiese & Fuhr (2007). Their compilations include VUV transition arrays $2s^22p^2 – [2s^22pn\ell$ ($n \leq 9$, 8 for $\ell = s$, $d$) + $2s2p^3$], extensive data for spin-allowed transitions $2s^22pn\ell – 2s^22pn'\ell'$ ($n' - n \geq 0$, $\ell' - \ell = \pm1$), and intercombination transitions $2s2p^3 – [2s^22pnp$ ($n \leq 5$) + $2s^22p4f$]. All these data have an evaluated uncertainty. We used their line strengths as reference values to evaluate uncertainties of the large amount of complementary data computed in the present work with Cowan's codes (Cowan 1981) in the LSF procedure. In general, the comparison was made for transitions not strongly affected by cancellations, i.e., those having the cancellation factor $|CF| \geq 0.1$. However, some transitions affected by cancellations were evaluated separately due to lack of dependable transition rates required for the intensity reduction method described in Section 5. The estimated averaged uncertainty was 10 %, 24 %, 40 %, and 52 % for transitions with computed line strengths $S_{ik} \geq 400$ atomic units (a.u.), $S_{ik} \in [150–400)$ a.u., [20–150) a.u., and [2–20) a.u., respectively. For $S_{ik} < 2$ a.u., the average uncertainty is about two orders of magnitude; nevertheless, some of those weak transitions were useful for the intensity reduction procedure. A few transitions affected by cancellations required for the intensity reduction were also found to be accurate within two orders of magnitude. In Table 2, the $A_{ki}$ values are supplemented with accuracy symbols and references to their sources. The accuracy code is explained in Table 10, and further details of the method applied here for the estimation of uncertainties can be found elsewhere (Kramida 2013a, 2013b, 2013c; Haris et al. 2014).

Table 10

Transition Probability Uncertainty Code

| Symbol | Uncertainty in *A*-value | Uncertainty in log(*gf*) |
|---|---|---|
| AAA | ≤0.3 % | ≤0.0013 |
| AA | ≤1 % | ≤0.004 |
| A+ | ≤2 % | ≤0.009 |
| A | ≤3 % | ≤0.013 |
| B+ | ≤7 % | ≤0.03 |
| B | ≤10 % | ≤0.04 |
| C+ | ≤18 % | ≤0.08 |
| C | ≤25 % | ≤0.11 |
| D+ | ≤40 % | ≤0.18 |
| D | ≤50 % | ≤0.24 |
| E | >50 % | >0.24 |

## 7. IONIZATION POTENTIAL

The ionization energy (IE) of C I was previously determined by Johansson (1966) to be 90 820.42(10) $cm^{-1}$. He used the $2pnp\ {}^3D_3$ ($n = 3–10$) series, which converged to the C II $2s^22p\ ^2P°_{3/2}$ limit at 90 883.84(10) cm$^{-1}$. Positions of these limits were later recalculated by Chang & Geller (1998) with their improved level values from solar IR measurements, yielding 90 883.854(15) cm$^{-1}$ for the C II $2s^22p\ ^2P°_{3/2}$ limit and IE = 90 820.469(15) cm$^{-1}$ after the subtraction of the C II $2s^22p\ ^2P°$ fine-structure interval accurately measured by Cooksy et al. (1986b). We noticed a small typo in the value of IE recommend by Chang & Geller (1998): with the C II $^2P°$ fine-structure interval of 63.395 09(2) cm$^{-1}$ from Cooksy et al. (1986b), it should be 0.010 cm$^{-1}$ lower than the value given by Chang & Geller.

As one of the results of the present work, more accurate level energies of $2pnp\ ^3D_3$ ($n$ = 3–10) and $2pnd\ ^3F°_4$ ($n$ = 3–6) are available now to derive the IE. Both series converge to C II $2s^22p\ ^2P°_{3/2}$. We used both the extended and modified Ritz quantum-defect expansions implemented in the RITZPL code by Sansonetti (2005). We modified this code by addition of a Monte Carlo module, which randomly varies the input level energies around their nominal values with a normal statistical distribution of width equal to the measurement uncertainty. Three-parameter exact fits of the extended and modified Ritz formulas (see Kramida 2013a) to the four-member $2pnd\ ^3F°_4$ ($n$ = 3–6) series yielded 90 883.83(24) cm$^{-1}$ and 90 883.86(24) cm$^{-1}$, respectively, where the specified uncertainties of the fit are entirely due to the measurement uncertainties of the levels. Three-parameter fits to the level values of the $2pnp\ ^3D_3$ ($n$ = 3–10) series gave limit values of 90 883.867(30) and 90 883.982(43) cm$^{-1}$ from the extended and modified three-term Ritz expansions, respectively, where the uncertainties are again dominated by the measurement uncertainties of the upper members of the series. A four-parameter fit of the same series diverges for the extended Ritz expansion, but yields a well-defined value of 90 883.862(72) cm$^{-1}$ for the modified expansion. The weighted mean of all five above values gives the C II $2s^22p\ ^2P°_{3/2}$ limit at 90 883.90(7) cm$^{-1}$, where the uncertainty is dominated by the mean deviation of individual determinations from the mean. Subtracting the C II $2s^22p\ ^2P°_{3/2}$ excitation energy (Cooksy et al. 1986b) yields the IE of C I to be 90 820.50(7) cm$^{-1}$. Comparing with the value 90 820.469(15) cm$^{-1}$ derived by Chang & Geller (1998) essentially from the same series data, one can see that the latter authors have grossly underestimated their uncertainty.

Another value of the IE was derived by Glab et al. (1998) using the $2pnp\ ^3D_3$ ($n$ = 35–70) Rydberg series measured with a VUV and UV double laser resonance technique. Glab et al. used a two-parameter fit of a simplified Rydberg formula $E_n = V - R_C(n - \delta)^{-2}$, where $E_n$ is the measured transition energy to the $2pnp\ ^3D_3$ level, $V$ is the IE from C I $2p3s\ ^3P°_2$ to C II $2s^22p\ ^2P°_{3/2}$, $R_C$ = 109 732.303 cm$^{-1}$ is the Rydberg constant for the carbon atom and $\delta$ is the quantum defect. The result of their fit was $V$ = 30 490.54(3) cm$^{-1}$, $\delta$ = 0.673(10), and the excitation energy of C I $2p3s\ ^3P°_2$ (60 393.14±0.05 cm$^{-1}$) was added to obtain the IE of C I, 90 820.33(8) cm$^{-1}$. We modified this value with the now precisely known energies of C I $2p3s\ ^3P°_2$ at 60 393.1693(14) cm$^{-1}$ (see Table 2) and C II $2s^22p\ ^2P°_{3/2}$ at 63.395 09(2) cm$^{-1}$ (Cooksy et al. 1986b). The corrected value is 90 820.31(3) cm$^{-1}$ with the uncertainty largely limited by the precision of wavelength measurements of Glab et al.

The latter authors have recently found that some of their measurements could be in error due to a wrong identification of the reference iodine molecule lines (Glab 2016). Subsequently, a re-analysis of those measurements has been made, which led to substantially improved accuracy. We have used those revised data for the high members of the $2pnp\ ^3D_3$ ($n$ = 40–69) series together with the low $n$ data ($n$ = 3–10) used in our preliminary determination and fitted this combined dataset with a four-term modified Ritz formula using the RITZPL code. The resulting final value of the IE of neutral carbon is 90 820.348(9) $cm^{-1}$ or 11.2602880(11) eV[1]. A detailed description of these revised measurements will be published elsewhere (see Glab et al. 2017).

## 8. CONCLUSION

The present work greatly extends the list of critically evaluated atomic-spectroscopy data on neutral carbon and improves its accuracy. The number of known energy levels of C I is extended from 282 (Kramida et al. 2016) to 412, and the list of critically evaluated transitions is extended from 1378 to 2102. Relative uncertainties of the new recommended energy levels vary between $10^{-9}$ and $5\times10^{-4}$. The list of critically evaluated transition probability data is extended by addition of 241 new values, increasing the total number of reliable *A*-values to 1616. In addition to the data for natural carbon, we provide comprehensive lists of energy level and wavelength data on isotopes $^{12}$C and $^{13}$C, which was derived by a combination of isotope shift measurements and calculations reported in the literature.

## ACKNOWLEDGMENTS

This work was partially funded by the Astrophysics Research and Analysis program of the National Aeronautics and Space Administration of the USA. Helpful discussions with Dr. Gillian Nave of NIST are gratefully acknowledged. She provided the raw data on the SiC FTS spectrum used in this work.

## References

Berengut, J. C., & Flambaum, V. V. 2010, HyInt, 196, 269
Berengut, J. C., Flambaum, V. V. & Kozlov M. G. 2006, PhRvA, 73, 012504
Bernheim, R. A., & Kittrell, C. 1980, AcSpe, 35, 51
Bowen, I. S., & Ingram, S. B. 1926, PhRv, 28, 444
Bowen, I. S. 1927, PhRv, 29, 231
Boyce, J. C. 1936, MNRAS, 96, 690
Boyce, J. C., & Robinson, H. A. 1936, JOSA, 26, 133

[1] Conversion of $cm^{-1}$ to eV unit was made with the use of their equivalence relation 1 eV = 8065.544005(50) $cm^{-1}$ as recommended by Mohr et al. (2016)

Braams, B. J., & Chung, H.-K. 2015, JPhCS, 576, 11001
Brault, J. W. 1978, in Future Solar Optical Observations Needs and Constraints, Proceedings of the JOSO Workshop, ed. G. Godoli, Vol. 106 (Firenze: Pubblicazioni della Universita degli Studi di Firenze), 33
Brault, J. W., & Abrams, M. C. 1989, Fourier Transform Spectroscopy: New Methods and Applications (1989 OSA Technical Digest Series, Vol 6; Santa Fe, NM Optical Society of America), 110
Burnett, C. R. 1950, PhRv, 80, 494
Cantù, A. M., Mazzoni, M., Pettini, M., & Tozzi, G. P. 1981, PhRvA, 23, 1223
Cardelli, J. A., Meyer, D. M., Jura, M., & Savage, B. D. 1996, ApJ, 467, 334
Chang, E. S., & Geller, M. 1998, PhyS, 58, 326
Clark, C. W. 1983, OExpr, 8, 572
Cooksy, A. L., Saykally, R. J., Brown, J. M., & Evenson, K. M. 1986a, ApJ, 309, 828
Cooksy, A. L., Blake, G. A., & Saykally, R. J. 1986b, ApJ, 305, L89
Cowan, R. D. 1981, The Theory of Atomic Structure and Spectra (Berkeley, CA: University of California Press) and Cowan code package for Windows by Kramida, A. available from http://das101.isan.troitsk.ru/COWAN
Curdt, W., Brekke, P., Feldman, U., et al. 2001, A&A, 375, 591
Curran, S. J., Tanna, A., Koch, F. E., et al. 2011, A&A, 533, A55
Edlén, B. 1933a, ZPhy, 84, 746
Edlén, B. 1933b, ZPhy, 85, 85
Edlén, B. 1934, Nov Acta Reg. Soc. Sci. Upsalien, Ser IV, 9, 1
Edlén, B. 1936, ZPhy, 98, 445
Edlén, B. 1947, Natur, 159, 129
Evans, T. L. 2010, JApA, 31, 177
Farmer, B., & Norton, H. 1989, A High-Resolution Atlas of the Infrared Spectrum of the Sun and the Earth Atmosphere from Space. A Compilation of ATMOS Spectra of the Region from 650 to 4800 cm-1 (2.3 to 16 µm), Vol. I., The Sun (NASA RP-1224; Washington: NASA)
Feldman, U., Brown, C. M., Doschek, G. A., Moore, C. E., & Rosenberg, F. D. 1976, JOSA, 66, 853
Feldman, U., & Doschek, G. A. 1991, ApJS, 75, 925
Fowler, A., & Selwyn, E. W. H. 1928, RSPSA, 118, 34
Frerking, M. A., Keene, J., Blake, G. A., & Phillips, T. G. 1989, ApJ, 344, 311
Froese Fischer, C. 2006, JPhB, 39, 2159
García-Hernández D. A., Hinkle K. H., Lambert D. L. & Eriksson K. 2009 ApJ 696 1733
Genzel, R., Harris, A. I., Stutzki, J., & Jaffe, D. T. 1988, ApJ, 332, 1049
Glab, W. L., Glynn, P. T., & Robicheaux, F. 1998, PhRvA, 58, 4014
Glab, W. L. 2016, private communication
Glab, W. L., Haris, K., & Kramida, A. 2017, Revision of the ionization energy of neutral carbon, to be published in JPhB
Goldbach, C., Martin, M., & Nollez, G. 1989, A&A, 221, 155
Goly, A. 1976, A&A, 52, 43
Grevesse, N. 1984, PhST, 8, 49
Griesmann, U., & Kling, R. 2000, ApJ, 536, L113
Griesmann, U., Kling, R., Burnett, J. H., & Bratasz, L. 1999, in Proc. SPIE 3818, ed. G. R. Carruthers, & K. F. Dymond, 180
Guelachvili, G., & Narahari Rao, K. 1986, Handbook of Infrared Standards I (Academic Press: Florida)
Haberstroh, R. A., Kossler, W. J., Ames, O., & Hamilton, D. R. 1964, PhRv, 136, B932
Haridass, C. & Huber, K. P. 1994, ApJ, 420, 433
Haris, K., Kramida, A., & Tauheed, A. 2014, PhyS, 89, 115403
Hase, F., Wallace, L., McLeod, S. D., Harrison, J. J., & Bernath, P. F. 2010, JQSRT, 111, 521
Henning, T., & Schnaiter, M. 1998, EM&P, 80, 179
Herzberg, G. 1958, RSPSA, 248, 309
Hibbert, A., Biémont, É., Godefroid, M., & Vaeck, N. 1993, A&AS, 99, 179
Hill, F., Bogart, R. S., Davey, A., et al. 2004, in Proc. SPIE 5493, Optimizing Scientific Return for Astronomy through Information Technologies, ed. P. J. Quinn, & A. Bridger (International Society for Optics and Photonics), 163
Hofmann, W., & Weissler, G. L. 1971, JOSA, 61, 223
Holmes, J. R. 1951, JOSA, 41, 360_1
Ingram, S. B. 1929, PhRv, 34, 421

Jaffe, D. T., Harris, A. I., Silber, M., Genzel, R., & Betz, A. L. 1985, ApJ, 290, L59
Johansson, L. 1966, Ark Fys, 31, 201
Johansson, L., & Litzén, U. 1965, Ark Fys, 29, 175
Johnson, R. C. 1925, RSPSA, 108, 343
Junkes, J., Salpeter, E. W., & Milazzo, G. 1965, Atomic Spectra in the Vacuum Ultraviolet from 2250 to 1100 Å Part One – Al, C, Cu, Fe, Ge, Hg, Si, ($H_2$), Specola Vaticana (Città del Vaticano)
Kaufman, V., & Ward, J. F. 1966b, JOSA, 56, 1591
Keenan, P. C., & Greenstein, J. L. 1963, The Line Spectrum of R Coronae Borealis, λλ 3700-8600 Å, (Contrib. from the Perkins Observatory), Vol. II, No:13
Keene, J., Schilke, P., Kooi, J., et al. 1998, ApJ, 494, L107
Kiess, C. C. 1938, JRNBS, 20, 33
Klein, C. H., Lewen, F., Schieder, R., Stutzki, J., & Winnewisser, G. 1998, ApJ, 494, 125
Knapp, G. R., Crosas, M., Young, K., & Ivezić, Ž. 2000, ApJ, 534, 324
Kozlov, M. G., Tupitsyn, I. I., & Reimers, D. 2009, PhRvA, 79, 022117
Kozlov, M. G. 2016 (private communication)
Kramida, A. E. 2011, CoPhC, 182, 419
Kramida, A. 2013a, Fusion Sci. Technol., 63, 313
Kramida, A. 2013b, NISTJ, 118, 168
Kramida, A. 2013c, NISTJ, 118, 52
Kramida, A., Ralchenko, Y., Reader, J., & NIST ASD Team, 2016, NIST Atomic Spectra Database, Version 5.4, (Gaithersburg, MD: National Institute of Standards and Technology), available at http://physics.nist.gov/asd
Labazan, I., Reinhold, E., Ubachs, W., & Flambaum, V. V. 2005, PhRvA, 71, 40501
Lambert, D. L., & Swings, J. P. 1967, SoPh, 2, 34
Langer, W. D. 2009, in Submillimeter Astrophysics and Technology: A Symposium Honoring Thomas G. Phillips ASP Conference Series, Vol. 417
Lee, N., & Edwards, A. K. 1975, PhRvA, 11, 1768
Levshakov, S. A., Combes, F., Boone, F., et al. 2012, A&A, 540, L9
Liu, X.-W., Barlow, M. J., Danziger, I. J., & Clegg, R. E. S. 1995, MNRAS, 273, 47
Livingston, W., & Wallace, L. 1991, An atlas of the solar spectrum in the infrared from 1850 to 9000 $cm^{-1}$ (1.1 to 5.4 μm), (NSO Technical Report, Tucson: National Solar Observatory)
Luo, D., & Pradhan, A. K. 1989, JPhB, 22, 3377
Maki, A. G., & Wells, J. S. 1992, NISTJ, 97, 409
Marrone, P. V., & Wurster, W. H. 1971, JQSRT, 11, 327
Mazzoni, M., Tozzi, G. P., Cantu', A. M., & Pettini, M. 1981, PhyBC, 111, 379
Meggers, W. F., & Humphreys, C. J. 1933, JRNBS, 10, 427
Meija, J., Coplen, T. B., Berglund, M., et al. 2016, PApCh, 88, 293
Merton, T. R., & Johnson, R. C. 1923, RSPSA, 103, 383
Minnhagen, L. 1954, Ark Fys, 7, 413
Minnhagen, L. 1958, Ark Fys, 14, 481
Mohr, P. J., Newell, D. B., & Taylor, B. N. 2016, JPCRD, 45, 043102
Moore, C. E. 1970, Selected Tables of Atomic Spectra: C I, C II, C III, C IV, C V, C VI, Nat. Stand. Ref. Data Ser., Nat. Bur. Stand., 3, Sec. 3, (Washington, DC: US Govt. Printing Office)
Moore, C. E. 1993, Tables of Spectra of Hydrogen, Carbon, Nitrogen, and Oxygen Atoms and Ions (Boca Raton, FL: CRC Press)
More, K. R., & Rieke, C. A. 1936, PhRv, 50, 1054
Murphy, M. T., & Berengut, J. C. 2014, MNRAS, 438, 388
Nave, G., Griesmann, U., Brault, J. W., & Abrams, M. C. 2015, XGREMLIN: Interferograms and spectra from Fourier transform spectrometers analysis, ascl, ascl:1511.004
Nussbaumer, H., & Storey, P. J. 1984, A&A, 140, 383
Parenti, S., Vial, J.-C., & Lemaire, P. 2005, A&A, 443, 679
Paschen, F., & Kruger, G. 1930, AnP, 399, 1
Peck, E. R., & Reeder, K. 1972, JOSA, 62, 958
Phillips, T. G., Huggins, P. J., Kuiper, T. B. H., & Miller, R. E. 1980, ApJ, 238, L103
Radziemski, L. J., & Andrew, K. L. 1965, JOSA, 55, 474
Redman, S. L., Nave, G., & Sansonetti, C. J. 2014, ApJS, 211, 4
Ryde, J. W. 1927, RSPSA, 117, 164

Saloman, E. B., & Sansonetti, C. J. 2004, JPCRD, 33, 1113
Sandlin, G. D., Bartoe, J.-D. F., Brueckner, G. E., Tousey, R., & Vanhoosier, M. E. 1986, ApJS, 61, 801
Sansonetti, C. J. 2005 (private communication)
Sansonetti, C. J. 2007, NISTJ, 112, 297
Saykally, R. J., & Evenson, K. M. 1980, ApJ, 238, L107
Shenstone, A. G. 1947, PhRv, 72, 411
Simeon, F. 1923, RSPSA, 102, 484
Sonzogni, A. 2016, NuDat v. 2.6 (National Nuclear Data Center, Brookhaven National Laboratory, Upton, NY, USA.) available at http://www.nndc.bnl.gov/nudat2/
Swensson, J. W. 1967, NW, 54, 440
Tachiev, G., & Fischer, C. F. 2001, CaJPh, 79, 955
Toon, G. C. 1991, OptPN, 2, 19
Ubaschs, W. 2016, private communication
Wilkinson, P. G. 1955, JOSA, 45, 862
Wilkinson, P. G., & Andrew, K. L. 1963, JOSA, 53, 710
Wallace, L., & Hinkle, K. 2007, ApJS, 169, 159
Wallace, L., Hinkle, K., & Livingston, W. C. 1993, An atlas of the photospheric spectrum from 8900 to 13600 $cm^{-1}$ (7350 to 11230 Å), (NSO Technical Report #93-001; Tucson: National Solar Observatory)
Wallace, L., Livingston, W., Hinkle, K., & Bernath, P. 1996, ApJS, 106, 165
Wang, Y., Zatsarinny, O., & Bartschat, K. 2013, PhRvA, 87, 12704
Whaling, W., Anderson, W. H. C., Carle, M. T., et al. 2002, NISTJ, 107, 149
Wiese, W. L., Fuhr, J. R., & Deters, T. M. 1996, Atomic Transition Probabilities of Carbon, Nitrogen, and Oxygen: A Critical Data Compilation, (JPCRD Monograph No. 7, New York: AIP)
Wiese, W. L., & Fuhr, J. R. 2007, JPCRD, 36, 1287; Erratum: 2007, 36, 1737
Wolber, G., Figger, H., Haberstroh, R. A., & Penselin, S. 1969, PhLA, 29, 461
Wolber, G., Figger, H., Haberstroh, R. A., & Penselin, S. 1970, ZPhy, 236, 337
Yamamoto, S., & Saito, S. 1991, ApJ, 370, L103